\DocumentMetadata{
  lang        = en-US,
  pdfversion  = 1.7,
}
\def\buildmode{full}
\documentclass[sigconf]{acmart}
\AtBeginDocument{%
  }

\newif\ifreporttotal

\ifdefined\appendixonly\providecommand{\buildmode}{appendix}\fi
\ifdefined\posteronly\providecommand{\buildmode}{main}\fi

\providecommand{\buildmode}{full}    

\newif\ifshowmain
\newif\ifshowappendix
\makeatletter
\def\bmd@full{full}
\def\bmd@main{main}
\def\bmd@appendix{appendix}
\ifx\buildmode\bmd@full
  \showmaintrue  \showappendixtrue
\else\ifx\buildmode\bmd@main
  \showmaintrue  \showappendixfalse
\else\ifx\buildmode\bmd@appendix
  \showmainfalse \showappendixtrue
\else
  \PackageError{buildmode}{Unknown \string\buildmode: `\buildmode'}%
    {Set \string\buildmode\space to full, main, or appendix.}%
\fi\fi\fi

\makeatletter
\ifshowmain\ifshowappendix
  \AtBeginDocument{%
    \@ifundefined{r@acm@mainpaperend}{}{%
      \edef\acm@mainpaperpages{\getpagerefnumber{acm@mainpaperend}}%
      \expandafter\xdef\csname r@TotPages\endcsname
        {{\acm@mainpaperpages}{\acm@mainpaperpages}{}{}{}}%
    }%
  }%
\fi\fi
\makeatother

\AtBeginDocument{%
  \ifshowmain\else
    \setcounter{table}{3}%
\newlabel{sec:intro}{{1}{1}{Introduction}{section.1}{}}
\newlabel{sec:related}{{2}{2}{Related Work}{section.2}{}}
\newlabel{sec:system}{{3}{2}{\lipcoder System}{section.3}{}}
\newlabel{sec:setup}{{4}{3}{Experimental Setup}{section.4}{}}
\newlabel{sec:results}{{5}{3}{Quantitative Results}{section.5}{}}
\newlabel{tab:study_outcomes_CL_multirow}{{1}{4}{Outcomes by task condition for visually impaired participants (\(n=5\)), reported as ``\emph {mean} ± \emph {standard deviation}'' (boldface indicates better results). SUS: System Usability Scale $\uparrow $ (0–100), TLX: NASA Task Load Index $\downarrow $ (0–100), Time $\downarrow $ (min), and Accuracy $\uparrow $ (\%)}{table.caption.2}{}}
\newlabel{tab:study_stats}{{2}{4}{Statistical significance test between the baseline and \lipcoder for visually impaired participants (\(n=5\)): $p$-values (paired t-test or Wilcoxon as appropriate), corrected $q$-values, and effect sizes (Cohen’s $d_z$). Shaded results indicate that \lipcoder showed better results}{table.caption.3}{}}
\newlabel{tab:feature_like}{{3}{4}{Perceived acceptance measures across \lipcoder features for visually impaired participants (\(n=5\)), on a 7-point Likert scale (1 = strongly disagree, 7 = strongly agree). PU: Perceived Usefulness $\uparrow $, PEOU: Perceived Ease of Use $\uparrow $, BI: Behavioral Intention to use $\uparrow $}{table.caption.4}{}}
\newlabel{sec:qualitative}{{6}{5}{Qualitative Analysis}{section.6}{}}
\newlabel{sec:discussion}{{7}{5}{Discussion and Conclusion}{section.7}{}}
{main-labels.tex not found; references into
        the main paper will print as ??}}%
  \fi
  \ifshowappendix\else
{appendix-labels.tex not found; references
        into the appendix will print as ??}}%
  \fi
}
\makeatother

\copyrightyear{2026}
\acmYear{2026}
\setcopyright{cc}
\setcctype{by}
\acmConference[ASSETS '26]{The 28th International ACM SIGACCESS Conference on Computers and Accessibility}{October 25--28, 2026}{Vila Nova de Gaia, Portugal}
\acmBooktitle{The 28th International ACM SIGACCESS Conference on Computers and Accessibility (ASSETS '26), October 25--28, 2026, Vila Nova de Gaia, Portugal}
\acmDOI{10.1145/3797867.3841242}
\acmISBN{979-8-4007-2521-0/2026/10}

\usepackage{kotex}
\usepackage{pifont}
\usepackage{xspace}
\usepackage{xcolor}
\usepackage{enumitem}
\usepackage{algorithm}
\usepackage{algpseudocode}
\usepackage{microtype}
\usepackage{array}
\usepackage{tabularx}
\usepackage{colortbl, xcolor}
\usepackage{afterpage}
\usepackage{balance}   
\usepackage{makecell}
\usepackage{booktabs, multirow, makecell, array}

\newcommand{\lipcoder}{\textsc{LipCoder}\xspace}

\newcommand{\resultnum}[2]{#1\small{ ±\,#2}}
\newcommand{\shade}{\cellcolor{teal!11}}

\makeatletter
\newcommand{\dashfill}{\leavevmode\cleaders\hbox{-}\hskip\fill\kern0pt}
\newcommand{\smalldashfill}{%
  \leavevmode
  \cleaders\hbox{$\cdotp$}\hskip\fill\kern0pt
}

\newcolumntype{C}[1]{>{\centering\arraybackslash}p{#1}}

\begin{document}

\title{\lipcoder: Voice-Enabled Coding Toolkit}

\author{Hayoon Kim}
\affiliation{%
  \institution{Music and Audio Research Group}
  \institution{Department of Intelligence and Information}
  \institution{Seoul National University}
  \city{Seoul}
  \country{Republic of Korea}}
\email{hyway@snu.ac.kr}

\author{Sungho Lee}
\affiliation{%
  \institution{Music and Audio Research Group}
  \institution{Department of Intelligence and Information}
  \institution{Seoul National University}
  \city{Seoul}
  \country{Republic of Korea}}
\email{sh-lee@snu.ac.kr}

\author{Juhwi Kim}
\affiliation{%
  \institution{Independent Researcher}
  \city{Seoul}
  \country{Republic of Korea}}
\email{juhwiedenkim@gmail.com}

\author{Bongwon Suh}
\affiliation{%
  \institution{Human-Centered Computing Lab}
  \institution{Department of Intelligence and Information}
  \institution{Seoul National University}
  \city{Seoul}
  \country{Republic of Korea}}
\email{bongwon@snu.ac.kr}

\author{Kyogu Lee}
\affiliation{%
  \institution{Music and Audio Research Group}
  \institution{Department of Intelligence and Information}
  \institution{AIIS}
  \institution{IPAI}
  \institution{Seoul National University}
  \city{Seoul}
  \country{Republic of Korea}}
\email{kglee@snu.ac.kr}

\makeatletter

\newcommand{\compactauthor}[3]{%
  \parbox[t]{0.30\textwidth}{%
    \centering
    {\@authorfont #1\textsuperscript{\(#2\)}\par}
    \vspace{0.15em}
    {\@affiliationfont Seoul, Republic of Korea\par}
    {\@affiliationfont #3\par}%
  }%
}

\renewcommand{\@mkauthors}{%
  \global\setbox\mktitle@bx=\vbox{%
    \noindent\unvbox\mktitle@bx\par
    \vskip 0.65em
    \centering

    \compactauthor{Hayoon Kim}{\flat\natural}{hyway@snu.ac.kr}\hfill
    \compactauthor{Sungho Lee}{\flat\natural}{sh-lee@snu.ac.kr}\hfill
    \compactauthor{Juhwi Kim}{\ast}{juhwiedenkim@gmail.com}\par

    \vskip 1.1em

    \compactauthor{Bongwon Suh}{\natural\ddagger}{bongwon@snu.ac.kr}%
    \hspace{0.08\textwidth}%
    \compactauthor{Kyogu Lee}{\flat\natural\sharp\dagger}{kglee@snu.ac.kr}\par

    \vskip 1.0em

    {\@affiliationfont
      \textsuperscript{\(\flat\)}
      Music and Audio Research Group
      \quad
      \textsuperscript{\(\ddagger\)}
      Human-Centered Computing Lab\par

      \textsuperscript{\(\natural\)}
      Department of Intelligence and Information, Seoul National University\par
      
      \textsuperscript{\(\sharp\)}
      AIIS, Seoul National University
      \quad
      \textsuperscript{\(\dagger\)}
      IPAI, Seoul National University
      \quad
      \textsuperscript{\(\ast\)}
      Independent Researcher\par
    }

    \vskip 1.0em
  }%
}
\makeatother

\renewcommand{\shortauthors}{Kim et al.}

\ifshowmain

\begin{abstract}
AI-assisted programming environments have accelerated software development, giving rise to new paradigms like vibe coding. However, their benefits remain largely inaccessible to visually impaired programmers, as existing screen readers and assistive tools offer limited support for these emerging workflows. We introduce \lipcoder, a voice-centric programming toolkit designed to deliver editor-level functionality through auditory and speech-based interfaces. \lipcoder offers features comprising speech feedback and earcon cues for comprehension and validation, as well as natural language input for navigation and modification. In an exploratory evaluation, 5 visually impaired programmers performed a series of coding tasks comparing \lipcoder with a baseline of VSCode, Copilot, and VoiceOver. Quantitative trends and qualitative feedback point to directions for auditory-first design that may broaden accessibility in speech-driven coding environments.


\end{abstract}




\maketitle

\else
  \twocolumn
  \thispagestyle{standardpagestyle}
\fi 


\ifshowmain
\section{Introduction}\label{sec:intro}

Modern integrated development environments (IDEs) \cite{vscode, visualstudio} offer syntax highlighting, structured navigation, live linting, and refactoring tools; LLM-powered assistants add a new paradigm of \emph{vibe coding} \cite{sarkar2025vibecoding} built on conversational interaction and context-aware code generation. Yet these benefits remain largely inaccessible to visually impaired (VI) programmers: most IDE features, including AI-driven ones, rely on visual channels that screen readers and braille displays cannot access efficiently \cite{sean2012exploratory, chen2025screenreaderusersvibe}. For example, a sighted programmer can glance at a GitHub Copilot \cite{copilot} suggestion and accept it in a fraction of a second, whereas a blind programmer must navigate auxiliary panels through a fragmented, multi-step process \cite{saviaga25impact}. Prior non-visual programming work---auditory debuggers, screen reader enhancements, tactile interfaces---restores individual cues but remains fragmented, and no existing system provides a unified, accessible AI-assisted coding experience \cite{potluri18codetalk, stefik09sodbeans, baker15structjumper}.

We introduce \lipcoder, a voice-centric programming toolkit that unifies modern IDE features into auditory and speech-driven interfaces (Figure~\ref{fig:teaser}, Appendix~\ref{app:lipcoder_usage_scenario}). \lipcoder pursues two goals: \emph{auditory accessibility}, bringing the glanceable efficiency of modern editors to auditory form so that blind and low-vision programmers can reach parity with sighted peers, and \emph{speech-driven vibe coding}, extending conversational coding workflows through speech. Implemented as a VSCode extension, \lipcoder integrates neural Automatic Speech Recognition (ASR), Text-To-Speech (TTS) with designed earcons, and an LLM-powered intent engine, in a modular design with swappable components.

We conducted an exploratory study with five visually impaired programmers, comparing \lipcoder against the real-world baseline (VSCode + Copilot + VoiceOver): participants showed numerically faster completion, lower perceived workload, and higher usability ratings, and interviews identified auditory summaries, earcons, and navigation as essential enablers of coding. A supplementary study with twenty sighted programmers is reported in Appendix~\ref{app:sighted_study}.

In short, we contribute (i) \lipcoder, a voice-centric programming toolkit unifying advanced auditory interaction with LLM-based vibe coding; (ii) a modular architecture (VSCode extension) whose LLM, TTS, and ASR components can be easily substituted for diverse accessibility needs; (iii) an exploratory evaluation with blind and low-vision programmers, reporting quantitative trends and qualitative insights against the real-world baseline.

\section{Related Work}\label{sec:related}
We organize prior work along two axes---modality and abstraction level---yielding four quadrants (Figure~\ref{fig:quadrants}, Appendix~\ref{app:related}). Traditional IDEs support navigation, comprehension, modification, and validation \cite{xia2018measuring}, but rely on color, spatial layout, and transient visual cues inaccessible to visually impaired developers. Audio-based toolkits add voice input (VoiceCode \cite{voicecode}, Dragon NaturallySpeaking \cite{dragon}) or auditory output---from screen-reader augmentations (CodeTalk \cite{potluri18codetalk}, StructJumper \cite{baker15structjumper}, IndentNav \cite{indentnav}) to non-visual IDEs (Emacspeak \cite{raman96emacspeak}, Sodbeans \cite{stefik09sodbeans})---targeting Discoverability, Glanceability, Navigability, and Alertability \cite{potluri18codetalk}, yet remain fragmented and impose steep learning curves. AI-assisted tools such as GitHub Copilot \cite{copilot}, Codex \cite{chen2021codex}, and Cursor \cite{cursor} assume visual workflows: suggestions appear on screen, and blind developers must navigate secondary lists and listen line by line \cite{saviaga25impact}. Hybrid voice+AI systems (Copilot Voice \cite{copilotvoice}, Serenade \cite{serenade}, Talon \cite{talonvoice}) couple speech input with AI generation yet still present output visually, forcing sequential screen-reader traversal to verify changes. No prior system unifies speech input, auditory output, and AI assistance within a single interaction loop---the gap \lipcoder{} targets; a tool comparison (Table~\ref{tab:related_work_comparison}) and the extended review appear in Appendix~\ref{app:related}.

\section{\lipcoder System}\label{sec:system}

\noindent\textbf{Design goals ---} \lipcoder unifies screen-reader and vibe-coding paradigms into an \emph{auditory-centric} model: integrating speech input, auditory output (speech and earcons), and an LLM-based intent engine enables both high-level code generation and fine-grained navigation without visual confirmation. Where existing systems fragment functionality across tools, \lipcoder combines accessibility features with vibe-coding principles, reducing context switching and pursuing auditory parity for blind programmers.

\noindent\textbf{Overview and implementation ---} \lipcoder is a Visual Studio Code extension \cite{vscode-extension} in TypeScript---VSCode was chosen for its cross-platform stability and rich extension ecosystem---built around an event-driven, modular core spanning three layers: IDE integration (edits and AST diffs via the Language Server Protocol), a core engine (state management, event dispatch, AI command parsing), and backend services (ASR, TTS, LLM APIs). A \emph{voice pipeline} (Whisper~\cite{radford2022whisper}/Silero~\cite{silero_models} with lightweight VAD) parses spoken commands at low latency; a \emph{syntax pipeline} supplies AST diffs for navigation and debugging; an \emph{audio scheduler} renders speech and earcons as concise, non-overlapping feedback. This decoupling makes backends hot-swappable---crucial for tailoring features to visually impaired developers' diverse needs; the architecture (Figure~\ref{fig:system}) is detailed in Appendix~\ref{app:implementation_details}.

\noindent\textbf{Features ---} \lipcoder provides 17 features in six categories, all invocable by ASR with speech or earcon feedback (full mapping in Table~\ref{tab:feature_mapping}, Appendix~\ref{app:feature_details}).

\noindent\textbf{Screen Reader and Project Overviews (F1--F6) ---} A custom screen reader adds timbral variation for syntax highlighting, earcon motifs for matching delimiters, stacked auditory intervals for indentation depth, and spatial panning of cursor position; spoken summaries additionally cover project structure and individual files---including images---without opening them.

\noindent\textbf{Navigation and Code Operations (F7--F13) ---} \lipcoder supports unit-wise navigation over symbols, functions, and syntax errors, plus natural-language code and interface navigation, via a hybrid command dictionary plus LLM-based intent classifier; LLM-based code generation and modification, code explanation, code execution, and log/error summarization are each confirmed with a spoken summary and an earcon.

\noindent\textbf{Alerts, Suggestions, and Control (F14--F17) ---} Pressing Enter triggers a minor earcon for syntax errors or a major earcon for clean lines; pause-triggered completions and a continuous, voice-navigable suggestion list surface LLM suggestions; ``Stop'' immediately halts ongoing TTS and features, and settings and help offer on-demand adjustment (e.g., TTS speed) and a spoken command cheat sheet. A usage scenario illustrating a realistic non-visual workflow appears in Appendix~\ref{app:lipcoder_usage_scenario}.

\section{Experimental Setup}\label{sec:setup}

\noindent\textbf{Participants ---} We conducted an exploratory, formative evaluation with 5 programmers with visual impairments (4 male, 1 female; age 17--50; M=12 years Python experience; at least 1 year of Python), recruited via local accessibility networks. We frame the study as formative given the difficulty of recruiting blind and low-vision developers; demographics appear in Table~\ref{tab:vip-demographics} (Appendix~\ref{app:study_details}). As a comparison group, we additionally ran the identical protocol with 20 sighted programmers (11 male, 9 female; age 21--31; M=3.7 years Python experience; at least 2 years of Python), recruited from a university programming community (Appendix~\ref{app:sighted_study}).

\noindent\textbf{Conditions ---} We adopted a within-subjects design with two conditions: \emph{Baseline} (VSCode + Copilot + VoiceOver) and \emph{\lipcoder} (VSCode + \lipcoder). The baseline represents the prevailing state of practice among blind and low-vision developers at the time of the study (mid-2025).

\noindent\textbf{Procedure ---} After informed consent and a demographic pre-survey, participants received training on both systems, then performed three task types (basic, boost, debugging) in each condition (six tasks, 15-minute cap); pilot-verified A/B task versions were counterbalanced, with task-type order fixed and system order randomized. After each task, participants completed SUS and NASA-TLX surveys, followed by a final feature acceptance (TAM) survey and a semi-structured interview. Task design was grounded in the navigation--comprehension--modification--validation framework of programming activity~\cite{xia2018measuring, minelli2014quantifying}; Figure~\ref{fig:procedure} and full details appear in Appendix~\ref{app:study_details}.

\noindent\textbf{Tasks ---} \emph{Basic} tasks required solving introductory problems 
using only Python's standard library, starting from function stubs raising \texttt{NotImplementedError}. \emph{Boost} tasks required external libraries for data analysis and plotting over JSON and CSV datasets with \texttt{matplotlib}. \emph{Debugging} tasks required locating and fixing three syntax and two logic errors in pre-written code.

\setcounter{table}{2}
\begin{table*}[!t]
\setlength{\tabcolsep}{2.5pt}
\renewcommand{\arraystretch}{1.2}
\caption{Perceived acceptance measures across \lipcoder features for visually impaired participants (\(n=5\)), on a 7-point Likert scale (1 = strongly disagree, 7 = strongly agree).
  PU: Perceived Usefulness $\uparrow$, PEOU: Perceived Ease of Use $\uparrow$, BI: Behavioral Intention to use $\uparrow$.}
\label{tab:feature_like}
\definecolor{BlueBase}{HTML}{97C0F6}
\definecolor{RedBase}{HTML}{EB5757}
\small
\begin{tabular}{clccccccccccccccccc}
\toprule
& Metric & F1 & F2 & F3 & F4 & F5 & F6 & F7 & F8 & F9 & F10 & F11 & F12 & F13 & F14 & F15 & F16 & F17 \\
\hline

\multirow{3}{*}{\rotatebox{90}{VI}} & PU $\uparrow$ & \cellcolor{RedBase!20} 3.40 & \cellcolor{BlueBase!40} 5.20 & \cellcolor{BlueBase!7} 4.20 & \cellcolor{BlueBase!60} 5.80 & \cellcolor{BlueBase!87} 6.60 & \cellcolor{BlueBase!80} 6.40 & \cellcolor{BlueBase!80} 6.40 & \cellcolor{BlueBase!73} 6.20 & \cellcolor{BlueBase!73} 6.20 & \cellcolor{BlueBase!100} 7.00 & \cellcolor{BlueBase!93} 6.80 & \cellcolor{BlueBase!93} 6.80 & \cellcolor{BlueBase!93} 6.80 & \cellcolor{BlueBase!80} 6.40 & \cellcolor{BlueBase!87} 6.60 & \cellcolor{BlueBase!100} 7.00 & \cellcolor{BlueBase!100} 7.00 \\
 & PEOU $\uparrow$ & \cellcolor{BlueBase!20} 4.60 & \cellcolor{BlueBase!20} 4.60 & \cellcolor{BlueBase!80} 6.40 & \cellcolor{BlueBase!87} 6.60 & \cellcolor{BlueBase!73} 6.20 & \cellcolor{BlueBase!93} 6.80 & \cellcolor{BlueBase!80} 6.40 & \cellcolor{BlueBase!80} 6.40 & \cellcolor{BlueBase!87} 6.60 & \cellcolor{BlueBase!87} 6.60 & \cellcolor{BlueBase!80} 6.40 & \cellcolor{BlueBase!87} 6.60 & \cellcolor{BlueBase!67} 6.00 & \cellcolor{BlueBase!80} 6.40 & \cellcolor{BlueBase!87} 6.60 & \cellcolor{BlueBase!100} 7.00 & \cellcolor{BlueBase!100} 7.00 \\
 & BI $\uparrow$ & \cellcolor{RedBase!20} 3.40 & \cellcolor{BlueBase!47} 5.40 & \cellcolor{BlueBase!20} 4.60 & \cellcolor{BlueBase!40} 5.20 & \cellcolor{BlueBase!93} 6.80 & \cellcolor{BlueBase!80} 6.40 & \cellcolor{BlueBase!67} 6.00 & \cellcolor{BlueBase!87} 6.60 & \cellcolor{BlueBase!73} 6.20 & \cellcolor{BlueBase!93} 6.80 & \cellcolor{BlueBase!100} 7.00 & \cellcolor{BlueBase!100} 7.00 & \cellcolor{BlueBase!93} 6.80 & \cellcolor{BlueBase!80} 6.40 & \cellcolor{BlueBase!100} 7.00 & \cellcolor{BlueBase!100} 7.00 & \cellcolor{BlueBase!93} 6.80 \\

\bottomrule
\end{tabular}
\end{table*}
\setcounter{table}{0}
\begin{table}[t]
  \centering
  \setlength{\tabcolsep}{3pt}
  \renewcommand{\arraystretch}{1.1}
  \caption{Outcomes by task condition for visually impaired participants (\(n=5\)), reported as \emph{mean} ± \emph{standard deviation}.
  SUS: System Usability Scale $\uparrow$ (0–100), TLX: NASA Task Load Index $\downarrow$ (0–100), Time $\downarrow$ (min), and
  Accuracy $\uparrow$ (\%).}
  \label{tab:study_outcomes_CL_multirow}
  \small
  \begin{tabular*}{\columnwidth}{@{\extracolsep{\fill}}l l *{4}{c}@{}}
    \toprule
    \multirow{2}{*}[-2pt]{\textbf{Task}} & \multirow{2}{*}[-2pt]{\textbf{System}} &
      \multicolumn{4}{c}{\textbf{Visually Impaired} (\(n=5\))} \\
    \cmidrule{3-6}
    & & SUS $\uparrow$ & TLX $\downarrow$ & Time $\downarrow$ & Acc. $\uparrow$ \\
    \midrule
    \multirow{2}{*}{\textbf{Basic}}
      & Baseline & \resultnum{37.0}{12.2} & \resultnum{61.7}{25.1} & \resultnum{13.1}{4.2} & \resultnum{52.0}{48.2} \\
      & \shade \lipcoder & \shade \textbf{\resultnum{64.5}{13.6}} & \shade \textbf{\resultnum{33.9}{12.3}} & \shade \textbf{\resultnum{10.9}{4.0}} & \shade \textbf{\resultnum{96.0}{8.9}} \\
    \midrule
    \multirow{2}{*}{\textbf{Boost}}
      & Baseline & \resultnum{41.5}{17.0} & \resultnum{42.2}{20.2} & \resultnum{12.9}{2.1} & \textbf{\resultnum{100.0}{0.0}} \\
      & \shade \lipcoder & \shade \textbf{\resultnum{70.5}{14.7}} & \shade \textbf{\resultnum{29.4}{12.8}} & \shade \textbf{\resultnum{10.7}{4.3}} & \shade \textbf{\resultnum{100.0}{0.0}} \\
    \midrule
    \multirow{2}{*}{\textbf{Debug}}
      & Baseline & \resultnum{45.0}{15.4} & \resultnum{36.7}{21.5} & \textbf{\resultnum{7.1}{3.8}} & \textbf{\resultnum{100.0}{0.0}} \\
      & \shade \lipcoder & \shade \textbf{\resultnum{71.5}{17.0}} & \shade \textbf{\resultnum{26.1}{10.1}} & \shade \resultnum{7.6}{5.0} & \shade \textbf{\resultnum{100.0}{0.0}} \\
    \midrule
    \multirow{2}{*}{\textbf{Overall}}
      & Baseline & \resultnum{41.2}{14.3} & \resultnum{46.9}{23.5} & \resultnum{11.0}{4.3} & \resultnum{84.0}{34.8} \\
      & \shade \lipcoder & \shade \textbf{\resultnum{68.8}{14.4}} & \shade \textbf{\resultnum{29.8}{11.4}} & \shade \textbf{\resultnum{9.8}{4.4}} & \shade \textbf{\resultnum{98.7}{5.2}} \\
    \bottomrule
  \end{tabular*}

\ifreporttotal
\vspace{5mm}
  
\begin{tabular}{l l c  *{5}{c}}
    \toprule
    \multirow{2}{*}[-2pt]{\textbf{Task}} & \multirow{2}{*}[-2pt]{\textbf{System}} & &
      \multicolumn{5}{c}{\textbf{Total} (\(n=25\))}
     
    \\
    \cmidrule{4-8}
    & & & SUS $\uparrow$ & TLX $\downarrow$ & Time $\downarrow$ & Acc. $\uparrow$ & \# Keys $\downarrow$ \\
    \midrule
    \multirow{2}{*}{\textbf{Base}}
      & Baseline & & \resultnum{53.5}{16.1} & \resultnum{48.7}{16.3} & \resultnum{14.0}{2.6} & \resultnum{56.0}{35.6} & -- \\
      & \shade \lipcoder & \shade & \shade \textbf{\resultnum{63.7}{13.1}} & \shade \textbf{\resultnum{39.8}{14.1}} & \shade \textbf{\resultnum{13.5}{2.4}} & \shade \textbf{\resultnum{84.8}{26.0}} & \shade -- \\
    \midrule
    \multirow{2}{*}{\textbf{Boost}}
      & Baseline & & \resultnum{58.4}{18.4} & \resultnum{41.6}{16.3} & \textbf{\resultnum{12.8}{3.3}} & \textbf{\resultnum{80.0}{31.6}} & -- \\
      & \shade \lipcoder & \shade & \shade \textbf{\resultnum{66.4}{14.6}} & \shade \textbf{\resultnum{38.5}{14.8}} & \shade \resultnum{13.4}{3.0} & \shade \resultnum{76.8}{28.7} & \shade -- \\
    \midrule
    \multirow{2}{*}{\textbf{Debug}}
      & Baseline & & \resultnum{63.2}{17.1} & \resultnum{35.1}{15.1} & \resultnum{11.2}{4.6} & \resultnum{92.8}{9.8} & -- \\
      & \shade \lipcoder & \shade & \shade \textbf{\resultnum{71.3}{13.0}} & \shade \textbf{\resultnum{32.8}{12.8}} & \shade \textbf{\resultnum{10.6}{5.0}} & \shade \resultnum{92.8}{9.8} & \shade -- \\
    \midrule
    \multirow{2}{*}{\textbf{Overall}}
      & Baseline & & \resultnum{58.4}{17.5} & \resultnum{41.8}{16.7} & \resultnum{12.7}{3.7} & \resultnum{76.3}{31.7} & -- \\
      & \shade \lipcoder & \shade & \shade \textbf{\resultnum{67.1}{13.8}} & \shade \textbf{\resultnum{37.0}{14.1}} & \shade \textbf{\resultnum{12.5}{3.8}} & \shade \textbf{\resultnum{84.8}{23.7}} & \shade -- \\
    \bottomrule
\end{tabular}
\fi
\end{table}     

\begin{table}[t]
  \centering
  \setlength{\tabcolsep}{1.5pt}
  \renewcommand{\arraystretch}{1.1}
  \caption{Statistical significance test between the baseline and \lipcoder for visually impaired participants (\(n=5\)): $p$-values (paired t-test or Wilcoxon as appropriate), corrected $q$-values, and effect sizes (Cohen’s $d_z$).}
  \label{tab:study_stats}
  \small
  \begin{tabular*}{\columnwidth}{@{\extracolsep{\fill}}
    l c
    *{3}{c} c
    *{3}{c} c
    *{3}{c} c
    *{3}{c}@{}
  }
    \toprule
    \multirow{2}{*}[-2pt]{\textbf{Task}} & &
      \multicolumn{3}{c}{\textbf{SUS}} & &
      \multicolumn{3}{c}{\textbf{TLX}} & &
      \multicolumn{3}{c}{\textbf{Time}} & &
      \multicolumn{3}{c}{\textbf{Acc.}}  \\
    \cmidrule{3-5}\cmidrule{7-9}\cmidrule{11-13}\cmidrule{15-17}
    & & p & q & $d_z$ & &
            p & q & $d_z$ & &
            p & q & $d_z$ & &
            p & q & $d_z$ \\
    \midrule
    \textbf{Basic}
     &
       & \shade 0.06 & \shade 0.27 & \shade 4.49
       & & \shade 0.13 & \shade 0.40 & \shade -1.15
       & & \shade 0.59 & \shade 0.76 & \shade -0.30
       & & \shade 0.19 & \shade 0.42 & \shade 0.82 \\
    \midrule
    \textbf{Boost}
     &
       & \shade 0.06 & \shade 0.27 & \shade 5.59
       & & \shade 0.13 & \shade 0.40 & \shade -1.08
       & & \shade 0.28 & \shade 0.52 & \shade -0.60
       & & 1.00 & 1.00 & 0.00 \\
    \midrule
    \textbf{Debug}
     &
       & \shade 0.06 & \shade 0.27 & \shade 6.33
       & & \shade 0.19 & \shade 0.42 & \shade -0.78
       & & 0.63 & 0.77 & 0.09
       & & 1.00 & 1.00 & 0.00 \\
    \midrule
    \textbf{Overall}
     &
       & \shade 0.06 & \shade 0.27 & \shade 10.15
       & & \shade 0.13 & \shade 0.40 & \shade -1.37
       & & \shade 0.63 & \shade 0.77 & \shade -0.26
       & & \shade 0.19 & \shade 0.42 & \shade 0.82 \\
    \bottomrule
  \end{tabular*}
\end{table}
\setcounter{table}{3}

\noindent\textbf{Metrics ---} We measured completion time, accuracy, SUS~\cite{brooke1996sus}, NASA-TLX~\cite{hart1988nasatlx}, per-feature TAM constructs~\cite{dabis1989perceived}, and interview transcripts; the rationale for these measures appears in Appendix~\ref{app:study_details}.

\section{Quantitative Results}\label{sec:results}

Overall, \lipcoder was associated with numerically favorable scores across nearly all metrics (Table~\ref{tab:study_outcomes_CL_multirow}); we report task completion time, NASA--TLX workload, SUS, and accuracy.

Mean completion times were numerically lower with \lipcoder in basic and boost tasks, though in debug tasks the baseline yielded a numerically lower mean ($7.1 \pm 3.8$ vs.\ $7.6 \pm 5.0$). NASA--TLX workload scores were numerically lower under \lipcoder in all tasks, and SUS ratings were numerically higher across all tasks (Overall: $41.2 \pm 14.3$ $\rightarrow$ $68.8 \pm 14.4$). Accuracy was at ceiling under both systems for boost and debug tasks, so it separates the systems only in basic tasks, where the baseline was low and highly variable (52.0 ± 48.2) against 96.0 ± 8.9 with \lipcoder.

\noindent\textbf{Significance ---} We conducted paired-sample $t$-tests \cite{ttest} or Wil\-cox\-on signed-rank tests \cite{wilcoxon} per Shapiro–\allowbreak Wilk nor\-mal\-ity \cite{shapiro}, with Holm–\allowbreak Bon\-fer\-roni cor\-rec\-tion \cite{holmbonferroni}; no comparison reached significance ($q > .05$ in all cases; Table~\ref{tab:study_stats}), so given the small sample ($n = 5$) the trends are descriptive; effect sizes are large for SUS ($d_z$ = 4.49--10.15) but unstable at this sample size (Appendix~\ref{app:extended_results}).

\noindent\textbf{Perceived acceptance ---} We measured acceptance of each \lipcoder feature via the TAM constructs Perceived Usefulness, Perceived Ease of Use, and Behavioral Intention \cite{dabis1989perceived} (Table~\ref{tab:feature_like}); maximum (7.0) ratings clustered on settings and help, continuous suggestions, code generation, code explanation, code execution, and pause-triggered suggestions.

\noindent\textbf{Sighted comparison group ---} The same directional trends held for the 20 sighted programmers but at a far smaller magnitude: SUS rose from $62.7 \pm 15.5$ to $66.7 \pm 13.7$ and NASA--TLX fell from $40.5 \pm 14.5$ to $38.8 \pm 14.1$, while completion time was unchanged ($13.1 \pm 3.5$ vs.\ $13.2 \pm 3.4$ min); again no comparison survived correction (Appendix~\ref{app:sighted_study}).

\section{Qualitative Analysis}\label{sec:qualitative}
Across the five VI participants, several consistent patterns emerged.

\noindent\textbf{Navigation, summarization, and generation ---} All participants relied on auditory navigation and summarization to reduce the overhead of line-by-line traversal: error and file-level summaries were described as ``essential,'' and natural-language navigation commands as ``game-changing.'' Speech-driven code generation was widely adopted, at granularities from function-by-function (P4) to whole-file (P5).

\noindent\textbf{Divisive and valued cues ---} Keyword voice differentiation was the most divisive feature: the low-vision P1 found tonal variation clarifying and fatigue-reducing, whereas P2--P5 found multiple voices redundant or distracting. Delimiter earcons and indentation cues were broadly valued (except by P5), and workflows adapted to residual vision and prior experience.

\noindent\textbf{Adoption needs ---} Participants wanted hybrid voice-plus-text input, configurability, and built-in tutorials for discoverability. P4 noted that Copilot with VoiceOver retains advantages for fine-grained review of AI suggestions. Two needs were more specific: P5, whose workplace bars LLM tools, valued speech-driven generation but worried about ceding control to AI-initiated restructuring, and P2 singled out spoken image captioning as what made tasks over charts and other visual assets workable. Full per-participant analyses appear in Appendix~\ref{app:per_participant}.

\noindent\textbf{Sighted programmers treat audio as augmentation ---} The comparison group inverted several of these priorities. They dismissed auditory output that duplicated what they could already see---voice differentiation and cursor panning were ``distracting,'' and several observed that ``reading is faster than listening''---but valued audio as a peripheral channel: delimiter and indentation earcons acted as error-prevention cues that made them ``snap back to attention,'' concise error and file summaries accelerated comprehension where verbatim readout overwhelmed, and progress sonification let them monitor long-running jobs away from the screen. Speech input was accepted for short commands yet rejected for sustained entry, since articulating evolving ideas aloud forces premature linearization and leaves little traceability of what changed. That the same cues were foundational for our visually impaired participants but merely supplementary here suggests auditory-first features transfer to sighted workflows only where they augment vision rather than replicate it (Appendix~\ref{app:sighted_study}).

\section{Discussion and Conclusion}\label{sec:discussion}

\noindent\textbf{Design Lessons ---}
Developing and evaluating \lipcoder yielded several lessons for voice-driven programming toolkits (elaborated in Appendix~\ref{app:extended_discussion}).
\emph{Divergent preferences:} participants preferred consistent concatenative synthesis over neural TTS, whose content-dependent prosody reads like text ``constantly changing its font.''
\emph{Leverage auditory strengths thoughtfully:} earcons and spatial panning aided comprehension, but effectiveness varied (panning helped low-vision but not congenitally blind users), so cues must be configurable.
\emph{Speech input streamlines navigation:} natural-language commands replace sequential traversal and complex shortcuts.
\emph{Voice commands are ambiguous:} \lipcoder prioritizes the current code context for underspecified requests; asking clarifying questions is future work.
\emph{Concise, coarse-to-fine summaries:} short explanations, expanded only on request, reduce working-memory load.
\emph{Tutorials matter:} comprehensive onboarding is critical for real-world adoption.
\emph{Unified yet modular:} integrating core features in one system reduces tool-juggling overhead, while swappable, cross-platform components serve diverse accessibility needs.

\noindent\textbf{Limitations ---}
\lipcoder does not yet cover all modern IDE features, such as build automation, testing, and version control.
Its performance also depends on ASR and LLM reliability and latency: misrecognition under noise or unfamiliar accents can cascade into misinterpreted intent, and high latency interrupts the coding flow.
Our study involved only five participants from a diverse population, and group dynamics and longitudinal use remain unstudied.

\noindent\textbf{Conclusion ---}
Our exploratory study suggests that auditory parity and speech-driven coding are feasible and valued by visually impaired programmers; while broader evaluations are needed, these findings highlight the potential of auditory-first toolkits for inclusive, AI-assisted development, and we open-source \lipcoder to foster continued research.

\fi


\ifshowmain
\begin{acks}
This work was partly supported by the research grant (No. 0767-20260032) from the Learning Sciences Research Institute at Seoul National University [50\%]; Youlchon Foundation (Nongshim Corporation and affiliated companies), Korea [45\%]; and by the Institute of Information \& communications Technology Planning \& Evaluation (IITP) grant (No. RS-2021-II211343), funded by the Korea government (MSIT) [5\%]. We thank Seungkyun Jeong, Donghyeon Han, Chanhong Kim, Sungjun Choi, Suvin Kim, Hyeonah Song, Hyongsop Kim, and Inho Seo for their support.
\end{acks}
\fi 


\bibliographystyle{ACM-Reference-Format}
\ifshowmain
  \ifshowappendix\else\balance\fi   
  \bibliography{refs}
  \ifshowappendix
    \label{acm@mainpaperend}%
    \appendix
\clearpage

\section{Extended Related Work}
\label{app:related}
We categorize related efforts along two dimensions: modality and accessibility (traditional visual vs. voice-based) and abstraction level (manual coding vs. AI-assisted coding) - yielding four quadrants as in Figure \ref{fig:quadrants}. In the following, we discuss prior work in each quadrant: traditional IDEs, voice programming interfaces, AI-assisted programming tools, and hybrid voice+AI approaches.

\subsection{$\mathcal{Q.A}$ - Traditional Programming Toolkits}
Modern IDEs and code editors provide a rich set of features that support four core programming activities: navigation, comprehension, modification, and validation \cite{xia2018measuring, cruz2017work, lawrance2013how}.
These affordances substantially improve coding efficiency by reducing the cognitive load of working with large and complex codebases.

\begin{enumerate}
    \item \textbf{Navigation.}
IDEs provide project-level navigation tools such as class and function outline views, project explorers, and call hierarchy panels.
These features allow sighted programmers to navigate large codebases efficiently and locate relevant code fragments or errors with minimal effort.

    \item \textbf{Comprehension.}
Visual code editing aids, including syntax highlighting, code folding, bracket matching, and minimaps, help developers quickly recognize syntactic categories, identify anomalies, and maintain awareness of structural hierarchies.
Such cues provide an at-a-glance understanding of program organization and project health.

    \item \textbf{Modification.}
Modern editors facilitate rapid code transformation through features such as inline refactorings, automated code generation (e.g., getters/setters), and context-sensitive suggestions.
These affordances reduce manual effort and streamline the editing process, supporting more efficient program evolution.

    \item \textbf{Validation.}
Continuous error detection and feedbacks, including real-time linting, inline alerts, autocompletion, and debugging visualizations enable a tight edit–diagnose–fix loop.
These mechanisms allow sighted developers to identify and resolve problems without disrupting the workflow.

\end{enumerate}
Although these capabilities define the expected baseline of modern programming environments, their reliance on color, spatial layout, and transient visual cues limits their utility to visually impaired developers.
Any novel programming interface, whether voice-driven or AI-driven, should therefore aim to provide accessible equivalents of these four fundamental functions to ensure developer productivity.

\begin{figure}
    \centering
    \includegraphics[width=\linewidth,alt={Two-by-two quadrant diagram: horizontal axis from unaccessible with voice to accessible with voice, vertical axis from low to high programming abstraction. Quadrant A bottom left is vision-centric utility coding, B bottom right is audio-centric utility coding, C top left is vision-centric vibe coding, and D top right is audio-centric vibe coding.}]{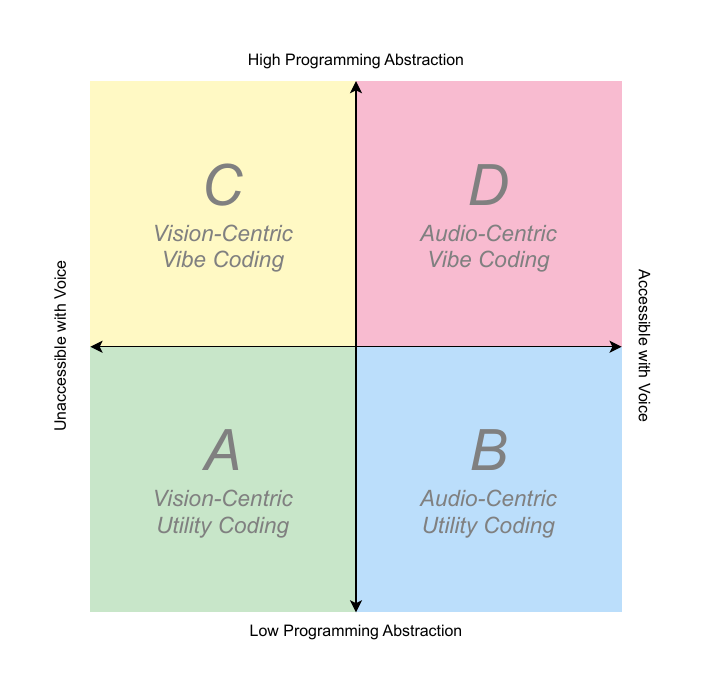}
    \caption{The quadrant model of programming paradigms. The $x$ axis corresponds to interaction mode (traditional visual vs. voice-based), and the $y$ axis corresponds to the abstraction level of coding (manual utility coding vs. AI-assisted coding).}
    \Description{A two-by-two quadrant diagram. The horizontal axis runs from "Unaccessible with Voice" on the left to "Accessible with Voice" on the right; the vertical axis runs from "Low Programming Abstraction" at the bottom to "High Programming Abstraction" at the top. Quadrant A, bottom left in green, is Vision-Centric Utility Coding. Quadrant B, bottom right in blue, is Audio-Centric Utility Coding. Quadrant C, top left in yellow, is Vision-Centric Vibe Coding. Quadrant D, top right in pink, is Audio-Centric Vibe Coding.}
    \label{fig:quadrants}
\end{figure}


\begin{table*}[t]
  \centering
  \setlength{\tabcolsep}{4pt}
  \renewcommand{\arraystretch}{1.15}
  \caption{Comparison of non-visual development environments across purposes, programming and accessibility aspects, and quadrant classification.}
  \label{tab:related_work_comparison}
  \resizebox{\textwidth}{!}{

  \begin{tabular}{p{5.2cm} p{6.8cm} ccc}
    \toprule
    \textbf{Environment \& Tool} & \textbf{Main purposes} & \textbf{Functionality} & \textbf{Accessibility} & \textbf{Q} \\
    \midrule
    StructJumper {\small (Emacs extension)}~\cite{baker15structjumper} 
      & Structure navigation, AST exploration 
      & Navigation & Navigability & A \\
    CodeTalk {\small (Visual Studio plugin)}~\cite{potluri18codetalk} 
      & Token reading \& summary, structure navigation 
      & Comprehension & Glanceability & B \\
    Tactile Code Skimmer {\small (Hardware)}~\cite{falase19tactile} 
      & Rendering of indentation profiles 
      & Comprehension & Glanceability & B \\
    Emacspeak {\small (Built into Emacs)}~\cite{raman96emacspeak} 
      & Menu \& buffer event narration 
      & Navigation & Navigability & B \\
    JavaSpeak {\small (Java IDE integration)}~\cite{smith00javaspeak} 
      & Java project navigation 
      & Navigation & Navigability & B \\
    Sodbeans {\small (NetBeans fork)}~\cite{stefik09sodbeans} 
      & GUI menu navigation, button/event announcements 
      & Navigation & Navigability & B \\
    Wicked Audio Debugger {\small (Standalone)}~\cite{stefik07wad} 
      & Control flow sonification; variable change timeline 
      & Validation & Alertability & B \\
    \midrule
    Copilot Voice {\small (VSCode plugin) \cite{copilotvoice}} 
      & Voice commands for code generation and  completions) 
      & Modification & Discoverability & D \\
    Serenade AI {\small (VSCode/Sublime plugin) \cite{serenade}} 
      & Voice-based code completion using AI 
      & Modification & Discoverability & D \\
    VoiceCode {\small (Atom/VSCode plugin) \cite{voicecode}}
      & Speech-input coding with AI-driven scaffolding 
      & Modification & Discoverability & D \\
    Talon Voice {\small (Voice-control engine) \cite{talonvoice}} 
      & Customizable voice-command framework for coding 
      & Modification & Navigability & B \\
    \midrule
    Baseline {\small (VSCode + Copilot + VoiceOver)} 
      & Editor + AI completion + screenreader
      & All & All & BD \\
    \textbf{LipCoder (ours)} {\small (VSCode extension)} 
      & Integrated features with ASR, TTS, and LLM 
      & All & All & BD \\
    \bottomrule
  \end{tabular}
  }
\end{table*}

\subsection{$\mathcal{Q.B}$ - Audio Based Programming Toolkits}

Audio-based programming toolkits can be broadly categorized into two types: systems that utilize audio as input and those that employ audio as output.

\subsubsection{\textbf{Voice as a Programming Input Modality}}
Programming by voice has been explored as an alternative to keyboard-and-mouse input, initially to improve accessibility, such as assisting programmers with motor disabilities or repetitive strain injuries (RSI), and more recently to enhance multitasking. Early systems such as VoiceCode \cite{voicecode} and Dragon NaturallySpeaking \cite{dragon} supported spoken commands for code snippets or editor actions, often through custom voice grammars for code entry, navigation, and editor control. However, these approaches were constrained by a limited set of predefined commands and provided predominantly visual feedback, limiting accessibility for blind and low-vision programmers.

\subsubsection{\textbf{Audio as a Programming Output Modality}}
For blind and low-vision programmers, the primary way to consume code is through screen readers or a braille display. They convert on-screen text and interface elements into synthesized speech or tactile interface.
There are OS-integrated screen readers such as Windows Narrator \cite{windows-narrator} and MacOS VoiceOver \cite{apple-voiceover}.
In practice, however, these built-in readers often fall short: their speech feedback can be fatiguing or inefficient to listen to, and they may not interact smoothly with all third-party software.
As such, visually impaired users more commonly use mainstream solutions such as NVDA \cite{nvda}, JAWS \cite{jaws}, and Sense Reader \cite{sense-reader}. However, vanilla screen readers were not initially optimized for programming, as they impose a high cognitive load due to verbose whitespace announcements, sequential navigation, and limited structural awareness.

To address these limitations, prior research has proposed programming specific augmentations, as addressed in Table \ref{tab:related_work_comparison}.
Potluri et. al \cite{potluri18codetalk} noted four key requirements for accessible programming: Discoverability, Glanceability, Navigability, and Alertability.

\begin{enumerate}
    \item \textbf{Discoverability.} The ease with which users can learn about available features and increase proficiency over time. Audio-based debuggers \cite{stefik07wad, stefik09sodbeans, potluri18codetalk} sonify control flow and variable changes to create an auditory execution timeline.
    The Wicked Audio Debugger \cite{stefik07wad} and Sodbeans' Sonified Omniscient Debugger \cite{stefik09sodbeans} sonify execution traces and variable updates, while CodeTalk's TalkPoints \cite{potluri18codetalk} allow setting auditory breakpoints that proactively announce runtime state. These approaches enable blind developers to construct and maintain mental models of program execution.

    \item \textbf{Glanceability.} The ability to quickly gain an overview of code or context without reading sequentially. Tactile interfaces \cite{falase19tactile} render indentation profiles through motorized sliders. Indentation-aware systems like IndentNav \cite{indentnav} and IndentTone \cite{indentone2016deprecated} as NVDA plugins replace redundant ``space'' narration with concise auditory cues or musical mappings. Other NVDA Coding plugins similarly improve workflows by offering shortcuts to jump across functions, announce indentation levels, and streamline navigation.

    \item \textbf{Navigability.} The efficiency with which users can move across code structures, files, or IDE panes. Multiple screen reader enhancements \cite{smith03nonvisual, baker15structjumper, potluri18codetalk, indentnav, indentone2016deprecated} expose abstract syntax tree (AST) hierarchies and announce IDE events via speech and earcons. Structural navigation features, such as Eclipse's accessible outline view \cite{smith03nonvisual} and StructJumper \cite{baker15structjumper}, enable rapid traversal of classes and methods. Semantic extensions, exemplified by CodeTalk \cite{potluri18codetalk}, provide higher-level navigation commands and event-driven alerts, while prototypes using earcons and spearcons convey control structures more efficiently.

    \item \textbf{Alertability.} The capacity of the environment to proactively notify users of important errors, events, or state changes. Non-visual IDEs such as Emacspeak \cite{raman96emacspeak}, JavaSpeak \cite{smith00javaspeak}, and Sodbeans \cite{stefik09sodbeans} integrate speech output, auditory menus, and debugging support natively, embedding alerts directly into the development environment.
\end{enumerate}

\subsection{$\mathcal{Q.C}$ - AI-Assisted Programming Toolkits}
In parallel with voice interfaces, the past few years have seen rapid growth in AI-assisted programming. Powered by LLMs, systems like GitHub Copilot \cite{copilot} and Codex \cite{chen2021codex} have introduced a ``conversational'' or context-aware coding paradigm. These tools can generate code suggestions or even entire functions based on a natural language prompt or the current code context. For example, Copilot can automatically suggest the following line or block of code as a developer types, and ChatGPT-like assistants can answer questions or produce code in response to instructions (e.g. ``Write a function to parse JSON''). By training on vast code corpora, such models often provide intelligent completions that save time on boilerplate and can even surface best practices.

AI coding assistants have been integrated into traditional development environments in various ways. Some, like Copilot in VS Code, appear as inline completions (grayed-out text that the developer can accept or ignore) or as pop-up suggestion lists. Others function as chatbots or side-panel assistants, where the developer can request code fixes or explanations. More advanced IDEs such as Cursor \cite{cursor} are emerging that tightly couple editing with an AI agent, allowing conversational refinement of code (the developer can say ``make this function asynchronous'' or ``explain this code'' in a chat and get immediate changes or answers). These AI-driven tools have shown productivity benefits in early studies, helping developers implement features faster and reducing the need to search for documentation \cite{bird2023taking}.

Despite their power, current AI programming aids assume a traditional visual workflow. The AI's output (generated code, error explanations, etc.) is typically presented on screen, relying on the developer to read and decide how to act on it. For sighted programmers, this works seamlessly; for example, a Copilot suggestion appears inline and can be accepted with a keystroke, making the interaction nearly frictionless. However, for a programmer who cannot easily see the suggestion, this process becomes cumbersome. As prior work \cite{saviaga25impact} has noted, blind developers using tools like Copilot do not have an immediate way to know what was suggested; they might only hear a sound cue and then must navigate into a secondary suggestions list, listen to the code read out line by line, and then confirm acceptance or rejection. In essence, AI-assisted coding to date has been optimized for sighted use, and when combined with assistive tech (like screen readers), it can disrupt workflow continuity. This limitation highlights an opportunity for multimodal AI programming interfaces that provide AI assistance through non-visual channels as well.

\subsection{$\mathcal{Q.D}$ - Hybrid Approaches: Audio + AI}
The fourth quadrant comprises systems that integrate voice interaction with AI code generation, forming conversational programming environments. These approaches aim to enable developers to express high-level intentions through speech, while an AI model generates or modifies the code accordingly, ideally with feedback delivered in an accessible manner. In principle, such systems combine the hands-free benefits of voice programming with the efficiency gains of AI automation.

Platforms like Copilot Voice \cite{copilotvoice} more tightly couple speech recognition with mainstream IDEs and AI assistants, offering inline completions, quick fixes, and contextual code generation triggered directly by spoken commands. Systems such as Serenade \cite{serenade} and Talon Voice \cite{talonvoice} extend this model by translating free-form utterances (e.g., ``create a for loop from 1 to 10'') into executable code edits via an AI backend. Together, these approaches reflect a broader trend toward vibe coding, where programming becomes more fluid, conversational, and intent-driven.

Despite these advances, most existing systems present their output primarily through visual channels: generated code appears inline, refactorings are marked with highlights, and errors are surfaced via tooltips or underlines. If screen reader and AI suggestions are not tightly integrated, blind and low vision developers must still traverse code sequentially with a screen reader to confirm modifications, fragmenting the workflow and undermining potential efficiency gains promised by speech- and AI-based interaction.

Thus, recent research emphasizes the need for holistic integration. A well-designed hybrid system should capture intent through speech, delegate execution to AI, and provide immediate auditory feedback on results - all within a single interaction loop. By unifying input, generation, and feedback, such environments can minimize context switching, reduce cognitive load, and approach auditory parity with the traditional visual IDE experience. Our work follows this trajectory by combining voice and AI assistance into one cohesive environment, addressing gaps in prior systems that treated the two components as loosely coupled.

\section{\lipcoder Usage Scenario}
\label{app:lipcoder_usage_scenario}
\begin{figure*}[t]
  \centering
  \includegraphics[width=\textwidth,alt={A programmer with closed eyes codes by voice: spoken commands such as Open training.py and Fix the syntax error and save pass through ASR into intent parsing and LipCoder actions, and spoken confirmations and earcons are returned through TTS while a monitor shows the code editor.}]{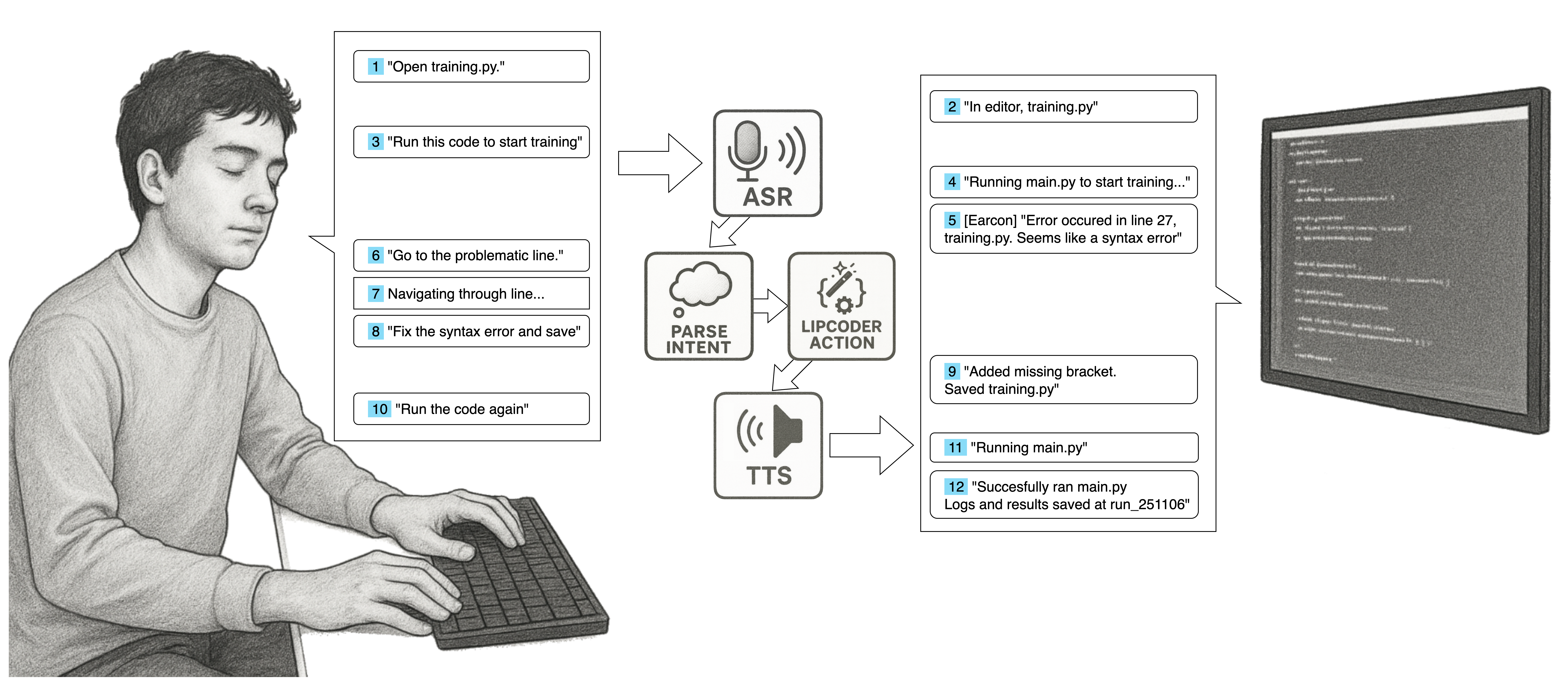}
  \caption{Overview of the \lipcoder\ system.}
  \Description{Illustration of a programmer with closed eyes coding by voice at a keyboard. Numbered speech bubbles show a spoken dialogue with LipCoder: the user says "Open training.py" and LipCoder replies "In editor, training.py"; the user says "Run this code to start training" and LipCoder replies "Running main.py to start training" and then plays an earcon with "Error occurred in line 27, training.py. Seems like a syntax error"; the user says "Go to the problematic line" and then "Fix the syntax error and save", and LipCoder replies "Added missing bracket. Saved training.py"; the user says "Run the code again" and LipCoder replies "Running main.py" and "Successfully ran main.py. Logs and results saved at run 251106". Arrows show that spoken commands pass through an ASR icon into Parse Intent and LipCoder Action blocks, and responses are rendered through a TTS icon; a monitor on the right displays the code editor.}
  \label{fig:teaser}
\end{figure*}

An overview of the \lipcoder system is shown in Figure~\ref{fig:teaser}.
To demonstrate how \lipcoder supports a realistic non-visual workflow, we describe the day-to-day experience of Flea, a visually impaired software engineer who relies on screen readers for development.

\paragraph{\textbf{Loading Workspace}}
Flea starts his work by opening the team's repository in VSCode.
As soon as the workspace loads, Flea says ``Project structure'' and \lipcoder announces the file tree, ``Seven source files, checkout.py, ..., utils.py, and whatever.py'' with soft earcons to mark each directory depth. This spoken overview allows Flea to mentally plan without scrolling through dozens of filenames. Flea navigates to the file, saying ``Open checkout.py.''

\paragraph{\textbf{Skimming the file}}
Flea now wants to get a quick mental model of the long checkout.py file before making any changes.
He issues the voice command, ``Symbol tree.''
\lipcoder speaks the top-level objects in order, ``6 classes, 21 functions, 3 constants,'' allowing him to navigate through the symbol structure.
He then speaks ``Function list.'' to grasp the names of functions.
This layered audio skimming provides Flea with the situational awareness that sighted developers gain from a minimap and outline panel.

\paragraph{\textbf{Locating the Bug}}
Flea runs the file by speaking ``Terminal (earcon) python checkout.py.'' \lipcoder voices out the log, ``TimeoutError in Line 106.''
He then issues a spoken command: ``Go to the function that raises TimeoutError.''
\lipcoder listens and parses the intent, jumps to the relevant function, then renders the first few lines in a timbre-coded voice—variables in a neutral woman, keywords in a lower-pitched man, etc.

\paragraph{\textbf{Quick Refactoring with Voice}}
Flea says, ``Wrap this try block in a with retry helper with three max attempts, and extract it as process\_retry.''
\lipcoder confirms, performs the extraction, and plays a success chime. The autogenerated helper appears, and \lipcoder immediately recites the new docstring summary so Flea can verify intent without skimming code verbosely.

\paragraph{\textbf{Writing a Test Function}}
Next, he says, ``Generate a pytest that asserts we attempt three retries on payment failure.''
\lipcoder instantly scaffolds the test, reads a crisp one-line summary, and parks the cursor inside the generated test function. Flea skims the snippet by nudging the cursor up and down; every line under the caret is spoken instantly. As he tweaks assertions, each keystroke is voiced, and the familiar Enter earcon provides continuous feedback- a bright major chord for syntactically sound lines and an immediate minor chord when syntax errors exist. This auditory safety net allows Flea to iterate fluidly, confident that he has not introduced a typo or an unbalanced bracket. When he pauses for more than 5 seconds, Earcon pops up to alert that completion suggestions exist. As Flea says ``Show suggestions.'', suggested code is played, aiding the user's fluent coding flow.

\paragraph{\textbf{High-level Comprehension Check}}
Before committing, Flea requests, ``Summarize changes.''
\lipcoder synthesizes: ``One new function process\_retry, three lines removed, eight lines added.''
Satisfied, he stages and pushes the patch.

\section{Feature Details}
\label{app:feature_details}
This appendix provides detailed descriptions of each \lipcoder feature category, including design rationale and implementation choices.
Table~\ref{tab:feature_mapping} maps every feature across modules, programming aspects, accessible programming aspects, and quadrant classification.
Some core features are inspired or derived from previous works and products. F2 (Earcons and matching delimiters) and F3 (Depth level earcons for structure hierarchy) are adapted from NVDA plugin. Some parts of F7 (Unitwise navigation) are derived from \cite{potluri18codetalk} and Xcode \cite{xcode}. F15 (Pause-triggered completion suggestions) is adapted from \cite{copilot}, and F16 (Continuous suggestions) is inspired from Perplexity \cite{perplexity}.
The quadrant letters denote: $\mathcal{A}$ corresponds to vision-centric utility coding, $\mathcal{B}$ to audio-centric utility coding, $\mathcal{C}$ to vision-centric vibe coding, and $\mathcal{D}$ to hybrid approaches that integrate audio-centric interaction with vibe coding toolkits.

\begin{table*}[t]
  \centering
  \caption{Mapping of LipCoder features across modules, programming aspects, accessible programming aspects, and quadrant classification. All features can be executed with ASR, text input, or shortcuts. AS refers to Audio scheduler. ASR could be an input for executing all commands, but could also be accessed by shortcuts and typing in the command palette.}
  \label{tab:feature_mapping}
  \setlength{\tabcolsep}{4pt}
  \renewcommand{\arraystretch}{1.15}
  \begin{tabular}{cllccc}
    \toprule
    \textbf{Label} & \textbf{Feature} & \textbf{Modules} & \textbf{Functionality} & \textbf{Accessibility} & \textbf{Q} \\
    \midrule
    \multicolumn{6}{l}{\it \textbf{Screenreader}} \\
      F1 & Diverse timbre vocal per token type & AS + TTS & Comprehension & Glanceability & B \\
      F2 & Earcons for matching delimiters & AS & Comprehension & Glanceability & B \\
      F3 & Depth level earcons for structure hierarchy & AS & Comprehension & Glanceability & B \\
      F4 & Spatial cursor panning & AS + TTS & Navigation & Glanceability & B \\
    \midrule
    \multicolumn{6}{l}{\it \textbf{Project Overview Readouts}} \\
      F5 & Project summarization & (ASR +) LLM + AS + TTS & Comprehension & Glanceability & D \\
      F6 & File explanation & (ASR +) LLM + AS & Comprehension & Glanceability & D \\
    \midrule
    \multicolumn{6}{l}{\it \textbf{Navigation \& Search}} \\
      F7 & Unitwise navigation & (ASR +) LLM + AS + TTS & Navigation & Navigability & B \\
      F8 & Natural-language code navigation & (ASR +) LLM + AS + TTS & Navigation & Navigability & D \\
      F9 & Natural-language interface navigation & (ASR +) LLM + AS + TTS & Navigation & Navigability & D \\
    \midrule
    \multicolumn{6}{l}{\it \textbf{Code Operations}} \\
      F10 & Code generation and modification & (ASR +) LLM + AS + TTS & Modification & Glanceability & D \\
      F11 & Code explanation & (ASR +) LLM + AS + TTS & Comprehension & Glanceability & D \\
      F12 & Code execution & (ASR +) LLM + AS + TTS & Validation & Alertability & B \\
      F13 & Log / Error summarization & (ASR +) LLM + AS + TTS & Validation & Alertability & D \\
    \midrule
    \multicolumn{6}{l}{\it \textbf{Alerts \& Suggestions}} \\
      F14 & Real-time syntax-error check & (ASR +) AS + TTS & Validation & Alertability & B \\
      F15 & Pause-triggered completion suggestions & (ASR +) LLM + AS + TTS & Modification & Alertability & D \\
      F16 & Continuous suggestions & (ASR +) LLM + AS + TTS & Modification & Alertability & D \\
    \midrule
    \multicolumn{6}{l}{\it \textbf{Control, Settings \& Help}} \\
      F17 & Settings and Help (cheat sheet) & (ASR +) LLM + AS + TTS & Comprehension & Discoverability & B \\
    \bottomrule
  \end{tabular}
\end{table*}

\subsection{Screen Reader (F1--F4)}
Unlike other features, the screen reader functions operate passively during every interaction with \lipcoder. During the initial design, we faced a critical decision: whether to make the system compatible with existing screen readers or to develop a custom screen reader from scratch. The first option offered clear advantages—visually impaired users could rely on familiar auditory outputs, thereby minimizing the learning curve. The second option, however, allowed us to experiment with novel auditory features and evaluate their potential to extend beyond the current limitations of mainstream screen readers. Because our development environment was MacOS, we adopted VoiceOver as the baseline system and studied its behavior. One notable drawback of VoiceOver is that it reads programming tokens literally, without adapting pronunciations to a coding context.

To explore alternatives, we experimented with a variety of open-source TTS engines, including Silero TTS \cite{silero_models}, Espeak-NG \cite{espeakng}, XTTS \cite{xtts}, and the native macOS VoiceOver voices. Pilot testing with five visually impaired programmers revealed a strong consensus: deep learning TTS systems were largely ineffective for code comprehension. Participants overwhelmingly preferred concatenative speech synthesis, especially VoiceOver, and several stated that neural voices were nearly impossible to recognize at all.

Based on these insights, we adopted a hybrid approach. We built our own screen reader that emulated VoiceOver interaction patterns, such as line-by-line and word-level reading, character echo, and cursor-based navigation, while retaining the most preferred VoiceOver female voice (Yuna) of Korean visually impaired VoiceOver users. On top of this familiar baseline, we implemented and evaluated several experimental auditory features:

\paragraph{Timbral variation for syntax highlighting (F1).} We explored whether timbral variation could serve as an auditory equivalent of syntax highlighting. Our implementation assigned distinct vocal timbres to code tokens, though in experimental settings we limited this to Python keywords, rendered using the male (Albert) voice, to reduce the initial learning burden.

\paragraph{Earcon motifs for delimiters (F2).} Inspired by NVDA, we introduced short earcon motifs to indicate opening and closing delimiters. For instance, `(' triggered a rising do--mi interval and `)' a descending mi--do interval. Pilot testing indicated that mapping more complex delimiters to longer auditory patterns was difficult to learn quickly, so in experiments, we restricted this feature to parentheses.

\paragraph{Depth level earcons (F3).} Also adapted from NVDA, we conveyed indentation depth using stacked auditory intervals. Successive indentations played do, do--mi, do--mi--sol, and do--mi--sol--si, while dedentation reversed the sequence. By contrast, native VoiceOver reports indentation numerically (e.g., ``four spaces''), which we considered less intuitive.

\paragraph{Spatial cursor panning (F4).} When reading a full line of code, we spatialized the auditory stream from left to right to provide a continuous sense of horizontal cursor position, giving programmers an embodied sense of code structure across the line.

Finally, because we had control over the implementation, we introduced concise auditory renderings for certain interface elements. For example, navigating a function list omitted verbose metadata such as line numbers or warnings, instead providing streamlined speech optimized for quick auditory scanning.

\subsection{Project Overview Readouts (F5--F6)}
\lipcoder maintains an internal model of the project directory tree, symbol table, and live syntax error list. Developers can invoke spoken summaries of this information at any time to obtain a high-level overview of the project structure (F5).

For finer-grained understanding, users can query individual files directly, for example, asking ``what does this file do?'' or asking specific file questions, even without opening the file (F6). This capability also extends to images---for instance, a user can ask about the content of an image. Although this functionality relies heavily on deep learning models, we implemented it to enhance accessibility for visually impaired developers.

\subsection{Semantic Navigation \& Search (F7--F9)}
Navigation represents one of the most significant burdens for visually impaired screen reader users. To reduce this burden, we integrated ASR-based commands that support navigation both within the code editor and across the development interface.

Inspired by CodeTalk \cite{potluri18codetalk} and grounded in VSCode's capabilities, we implemented unit-wise navigation (F7). Users can navigate through high-level program structures, including the symbol tree, function list, syntax errors, and breadcrumb units. For example, when a user says ``Show function list,'' a function overview is displayed; the user may then traverse the list with simple up/down commands and press Enter to place the cursor at the start of the chosen function.

Beyond unitwise navigation, we also support natural language navigation (F8). Users may issue commands such as ``Go to the top of the editor,'' ``Move to the definition of this function,'' ``Open snowball.js,'' or ``Navigate to whatever.py.'' Navigation is not restricted to code; we also enable interface-level navigation (F9), allowing users to switch contexts with commands like ``Go to terminal'' or ``Open settings.''

A core design challenge is linguistic variability: the same intent may be expressed in multiple ways (e.g., ``Go to variable x,'' ``Move to variable x,'' or even ``I would sincerely thank you if you could move my cursor to variable x''). To address this, we adopted a hybrid strategy. For speed, we curated a custom set of likely commands similar to Talon Voice \cite{talonvoice}, which we refer to as the command dictionary, matched to spoken input. For out-of-template utterances, we employ an LLM-based intent classifier to infer user intent and execute the appropriate command.

\subsection{Code Operations (F10--F13)}
Building on the generative capabilities of LLMs \cite{brown2020gpt3}, \lipcoder provides code generation (F10), code explanation (F11), and terminal log summarization (F13). Each operation is confirmed with a spoken summary and an earcon, allowing users to verify the change without leaving the audio workflow.

One can ask to modify or generate new code, and \lipcoder reads the peripheral context of the code to generate and modify appropriate code. When generated code comes from LLM, \lipcoder detects the changed parts and fits them inside the editor.

The user can either ask what a specific function or variable does, or just request the explanation of the full code in the editor. When one asks for an explanation of the terminal output, \lipcoder tracks the recent output and summarizes it through LLM. When an error occurs, \lipcoder explains the error and suggests a probable solution.

For simplicity, \lipcoder also provides a code execution feature (F12). A simple ``Execute this file'' command is mapped to the appropriate code execution command in the terminal (e.g., for a Python file, \texttt{python file.py}), which is guessed from the current project and file extension.

\subsection{Alerts \& Suggestions (F14--F16)}
\lipcoder tracks syntax errors to provide syntax error list navigation. In addition, pressing Enter triggers an on-the-fly LSP parse of the line just completed (F14); if a syntax error is detected, \lipcoder plays a minor `error-enter' earcon, whereas syntactically clean lines are confirmed with a major `normal-enter' earcon.

\lipcoder also offers pause-triggered completion suggestions as in Copilot \cite{copilot} (F15). When the user pauses typing for a few seconds and LLM-generated suggestions are available, \lipcoder plays a brief earcon. Pressing Shift+Enter will vocalize the top suggestion, allowing the developer to apply it or dismiss it.

In addition, \lipcoder offers continuous suggestions generated by LLM (F16). When the user issues ``Show suggestions,'' a suggestion list appears, which allows users to navigate through suggestions and pick them. Possible suggestions include ``Fix all syntax errors,'' ``Complete function x,'' ``Generate test functions for function x,'' etc.

\subsection{Control, Settings and Help (F17)}
A concise voice-first control layer keeps the auditory channel responsive. The command ``Stop'' immediately halts the ongoing TTS and the ongoing features being executed. An example is stopping code generation during its thinking process.

Additionally, settings and help (F17) allow adjustment of settings such as TTS playback speed on demand. Uttering ``Help'' triggers a brief spoken cheat sheet enumerating all available commands, ensuring discoverability without leaving the audio workflow.

\section{Implementation Details}
\label{app:implementation_details}

\begin{figure*}
    \centering
    \includegraphics[width=.98\linewidth,alt={Block diagram: voice, chat, code, and shortcut inputs flow through voice analysis with VAD and ASR, syntax analysis via LSP and AST, and intention analysis combining a command dictionary with an LLM fallback, into a central feature execution block, which fans out to UI interaction, UI rendering, LLM response generation, and audio scheduling that drives concatenative or neural speech synthesis to produce speech and earcon outputs.}]{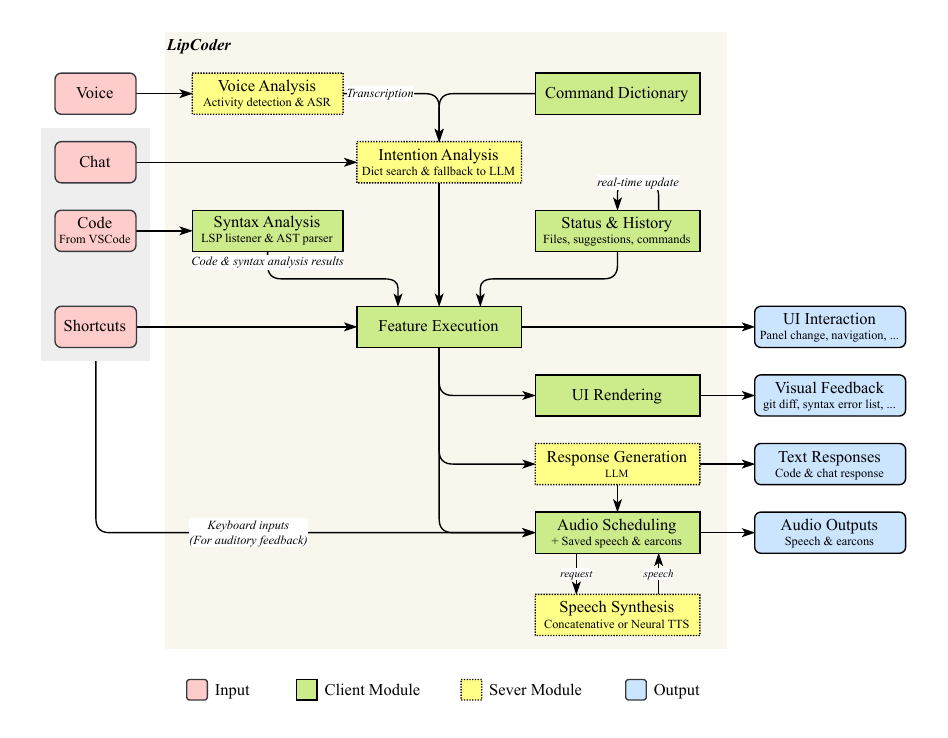}
    \caption{Implementation architecture of \lipcoder. 
    Some modules are executed on the server side, while others are on the client side; see the legend (bottom of the figure).
    \lipcoder supports four input types: voice, chat, code, and shortcuts. The latter three are keyboard inputs, and immediate auditory feedback is provided.
    The voice input is routed to the voice analysis stage, comprising Voice Activity Detection (VAD) and Automatic Speech Recognition (ASR) submodules, transcribing the speech into commands. The command text, either given directly by chat or transcribed from voice, is passed to the intention analysis module. 
    In this stage, a dictionary-based command search is first performed, then falls back to a Large Language Model (LLM) if the search gives no results.
    \lipcoder also maintains the current and past status of the project, including the files (and their changes), suggestions, and commands.
    These status data are passed for the feature execution, along with the code (preprocessed with the syntax analysis module), command, or shortcut inputs.
    The execution results in various outputs: interaction with UI, visual feedback in the VSCode UI, text responses (both in code and chat), and audio outputs. The text responses are generated with LLM, and audio outputs are controlled by the audio scheduling module, communicating with speech synthesis modules on the server side if necessary.
    Note that each \lipcoder feature executes only the necessary part of the entire system.
    }
    \Description{Block diagram of the LipCoder architecture. On the left, four pink input blocks: Voice, Chat, Code from VSCode, and Shortcuts. Voice feeds a Voice Analysis block (voice activity detection and ASR) whose transcription goes to Intention Analysis, a two-stage block combining dictionary-based command search with fallback to an LLM, supported by a Command Dictionary and a Status and History store of files, suggestions, and commands. Code feeds a Syntax Analysis block (LSP listener and AST parser). All streams converge on a central Feature Execution block. Downstream, Feature Execution fans out to UI Interaction (panel change and navigation), UI Rendering (visual feedback such as git diff and syntax error list), Response Generation (LLM-based text responses for code and chat), and Audio Scheduling (keyed speech and earcons), which drives Speech Synthesis (concatenative or neural TTS) to produce audio outputs of speech and earcons. A legend indicates pink blocks are inputs, green blocks are client modules, yellow blocks are server modules, and blue blocks are outputs.}
    \label{fig:system}
\end{figure*}


The system architecture of \lipcoder is shown in Figure~\ref{fig:system}.
\lipcoder is implemented as a VSCode extension \cite{vscode-extension} written in TypeScript. At its core is an event-driven message bus that links input modules, analyzers, feature executors, and output renderers. Instead of calling each other directly, modules publish and subscribe to typed events, which reduces latency and allows backends to be swapped seamlessly. Put simply, the pipeline captures inputs, analyzes them (e.g., by voice, syntax, or intent), executes the corresponding feature, and renders the result as visual, textual, or auditory output.

\begin{itemize}
    \item \textbf{Inputs.} Four input sources feed the system: Voice, Chat, Code (via VSCode), and user Shortcuts. Keyboard activity is also forwarded to the audio layer for auditory feedback.
    \item \textbf{Voice Analysis.} Audio activity detection and ASR run through configurable backends. By default, we use local services Whisper \cite{radford2022whisper} and Silero ASR, with a cloud fallback. Transcriptions are emitted on the bus and consumed by intention analysis. An on‑device Silero Voice Activity Detection (VAD) \cite{silero_models} front‑end gates audio before decoding and emits \textit{speech\_start}/\textit{speech\_end} events on the message bus. We use a dual‑threshold policy with max‑segment/timeouts for early endpointing; this reduces latency, cross‑talk, and hallucinations. When VAD is unavailable, we fall back to energy‑based VAD. Transcripts stream token‑by‑token to intention analysis; finalization occurs on the VAD end or timeout.
    \item \textbf{Syntax Analysis.} The extension subscribes to VSCode's LSP diagnostics and symbol streams for TypeScript/JavaScript and Python, and adapters expose a uniform AST/syntax event layer. These events inform navigation, refactoring, and auditory feedback.
    \item \textbf{Intention Analysis.} Transcripts and chat requests are first matched against a command dictionary; unmatched intents fall back to an external LLM for parsing and suggestion. A \textit{Status \& History} store tracks open files, recent commands, and suggestions to provide context and real‑time updates.
    \item \textbf{Feature Execution.} Resolved intents dispatch to features (navigation, edits, refactors, searches). Executions update the store and fan out to \textit{UI Rendering}, \textit{Response Generation}, and \textit{Audio Scheduling}.
    \item \textbf{UI Rendering and Feedback.} Using the VSCode Extension API, the client updates panels, selections, and decorations, and shows visual feedback such as diffs and diagnostic lists.
    \item \textbf{Response Generation.} For summarization, completion, and refactoring, the client calls a small Node/Python service that proxies to a configurable external LLM endpoint with streaming support.
    \item \textbf{Audio Scheduling.} A mixer queues and ducks synthesized speech with earcons and stored prompts. Prebuilt assets live in the client, and scheduling guarantees non‑overlapping playback with a low‑latency start.
    \item \textbf{Earcons.} Short, structured earcons convey code syntax and editor state. Events such as enter, parentheses/braces, and indentation are mapped to parameterized cues. Earcons are spatialized by nesting depth and paired with brief pitch contours to differentiate open versus closed.
    \item \textbf{Speech Synthesis.} TTS is pluggable; we currently support macOS VoiceOver, eSpeak‑NG, Silero TTS, and XTTSv2. Backends run locally behind the same event bus and stream PCM to the client, with dynamic fallback based on availability and latency.
\end{itemize}

\section{Study Details}
\label{app:study_details}
This appendix provides the details of the user study with visually impaired programmers: participant demographics (Table~\ref{tab:vip-demographics}), the experiment procedure (Figure~\ref{fig:procedure}), the rationale for the baseline and task design, full task descriptions, and the metric rationale.

\begin{table*}
    \caption{Demographics of visually impaired programmers. For each participant, screen readers, LLM/vibe coding tools, and primary programming languages are listed in order of preference (left to right). Abbreviations: SR = Sense Reader, VO = VoiceOver.}
    \label{tab:vip-demographics}
  \setlength{\tabcolsep}{3pt}
  \renewcommand{\arraystretch}{1.1}
  \resizebox{\textwidth}{!}{%
    \begin{tabular}{lllllllllllll}
        \toprule
        
        \multirow{2}{*}[-2pt]{\textbf{ID}} 
        & & \multicolumn{2}{c}{\textbf{General Info.}} 
        & & \multicolumn{3}{c}{\textbf{Development Setup}} 
        & & \multicolumn{3}{c}{\textbf{Coding Experience}} 
        \\
        \cmidrule{3-4}
        \cmidrule{6-8}
        \cmidrule{10-12}
        & & Type of VI & Gender & & Preferred IDE & Screen Reader(s) & LLM / Vibe Coding Tools & & Main Languages & Python & Total \\
        \midrule
        \bf P1 & & Low Vision & Female & & PyCharm & SR & ChatGPT & & Java, Python & 1 yr & 1.5 yrs \\
        \bf P2 & & Blind      & Male   & & Nano & VO, SR, NVDA & ChatGPT / Windsurf, Gemini-cli & & Python, JavaScript & 5 yrs & 6 yrs \\
        \bf P3 & & Low Vision & Male   & & VSCode & VO, SR, JAWS & Gemini / Windsurf & & Python, Swift & 20 yrs & 27 yrs \\
        \bf P4 & & Blind      & Male   & & VSCode & VO, SR, NVDA & ChatGPT, Gemini / Copilot & & JavaScript, Python & 1 yr & 6 yrs \\
        \bf P5 & & Blind      & Male   & & XCode & VO, SR, JAWS & ChatGPT & & Objective-C, Python & 6 yrs & 10 yrs \\
        \bottomrule
    \end{tabular}%
    }
\end{table*}
\begin{figure*}
    \centering
    \includegraphics[width=.73\linewidth,alt={Timeline in minutes: pre-survey, 15-minute training, then a 100-minute block repeated three times containing two 15-minute programming tasks each followed by a post-survey, ending with a final survey and a 15-minute interview.}]{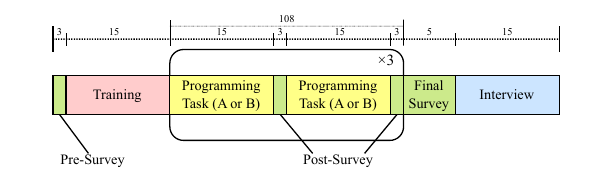}
    \caption{Experiment Procedure. Number units are minutes.}
    \Description{Horizontal timeline of the experiment session with durations in minutes. After a pre-survey, participants complete a 15-minute training session. Then a 100-minute block, repeated three times (once per task type), contains two 15-minute programming tasks (version A or B, one with each system), each followed by a post-survey. The session ends with a final survey and a 15-minute interview.}
    \label{fig:procedure}
\end{figure*}

\subsection{Baseline Choice}
We chose VSCode + Copilot + VoiceOver as our baseline, rather than systems such as Cursor or Serenade. At the time of our study (mid-2025), these alternatives had limited adoption among blind and low-vision developers and their accessibility support was still maturing. The combination of VSCode (the de facto standard editor), Copilot (the most widely deployed AI coding assistant), and VoiceOver (the default macOS screen reader) represented the prevailing state of practice among our target users. Using this setup as a baseline ensures that our evaluation compares against the tools that visually impaired programmers were actually relying on at the time, rather than against less representative configurations.

\subsection{Procedure}
The experiment procedure is illustrated in Figure \ref{fig:procedure}.
All participants were first informed about the study procedures and gave their consent to participate, including their agreement to interview recordings and logging of programming activities. The participants then completed a demographic pre-survey (Table~\ref{tab:vip-demographics}). After, they attended a training session on both the baseline environment and \lipcoder to familiarize themselves with the available features.
After the training session, participants performed three sets of tasks in each condition: basic, boost, and debugging, resulting in six programming tasks in total. To avoid learning effects, we created two parallel versions of each task (labeled A and B) and counterbalanced their assignment across conditions. Three pilot tests confirmed that versions A and B were similar in difficulty and expected completion time.
Each programming task was capped at 15 minutes, with time reminders given at 5-minute intervals.
Although the overall order of the task was fixed (basic, boost, debugging), the presentation order of \lipcoder and the baseline within each task was randomized, as was the assignment of programming question A or B.
After each programming task, participants completed the post-survey.
Finally, they conducted a final feature analysis survey, followed by a semi-structured interview to discuss their subjective impressions of the two systems.

\subsection{Task Design Rationale}
We designed simplified programming tasks that capture core dimensions of programming activity without requiring large-scale projects. While these toy problems cannot reflect the full complexity of professional development, they allow controlled comparisons across conditions and highlight where \lipcoder can make a difference.
Building on prior work \cite{minelli2014quantifying, xia2018measuring, cruz2017work}, we organized programming activity into four broad dimensions: navigation, comprehension, modification, and validation.
Our task design was informed by this framework.
Navigation and comprehension are unavoidable aspects of nearly all programming activities, and validation naturally occurs through code execution.
However, modification and validation can manifest in different ways. For example, modification may involve writing new code from scratch or altering existing code, while validation can appear either implicitly during development or explicitly in the form of systematic debugging.
To capture this spectrum, we separated debugging as a distinct task type, following previous studies that also contrasted general programming tasks with debugging scenarios \cite{lawrance2013how}.

Furthermore, we divided the programming tasks into two levels, \emph{basic} and \emph{boost}.
The basic tasks involved only the Python standard library, ensuring accessibility and comparability among participants.
In contrast, the boost tasks required the use of external libraries and the generation of visual output, such as plots.
This distinction allowed us to examine how participants adapted their workflows in familiar versus less familiar coding contexts.
We anticipated that voice- and AI-assisted programming might reveal different patterns of adoption and effectiveness depending on whether the task required well-known routines or novel, higher-level integrations.
This tiered structure ensured coverage of both low-level procedural skills and more complex exploratory activities, while maintaining the overall tractability of the experiment.

\subsection{Tasks}
\subsubsection{\textbf{Basic Task.}}
Basic tasks required participants to solve introductory programming problems using only Python's standard library (e.g., string or list manipulations). These tasks were designed to be straightforward and accessible without external dependencies. The first task implemented a \texttt{BirthdayCalculator} class to calculate the age of a given date of birth, following previous work \cite{saviaga25impact, kazemitabaar2023studying}. The second task implemented a \texttt{PasswordStrengthChecker} class to assess the validity of the password using only the built-in modules. For both tasks, participants received function stubs containing \texttt{NotImplementedError}.

\subsubsection{\textbf{Boost Task.}}
Boost tasks involved more advanced assignments that required the import and use of external libraries for data analysis and visualization. These tasks assessed the participants' ability to integrate third-party packages, manipulate structured data, and generate meaningful plots.

The first task focused on a JSON dataset of U.S. universities. The participants implemented five functions: (1) loading the data, (2) counting the universities by state, (3) identifying the top-$N$ states, (4) preparing data for visualization, and (5) generating a bar chart.

The second task centered on a CSV dataset of movie records. The participants implemented five analogous functions: (1) loading the dataset into a list of movies, (2) computing the number of movies per year, (3) calculating genre distributions by year, (4) preparing data for plotting, and (5) generating a line plot of genre counts over time.

As in the basic tasks, each function was initially stubbed with \texttt{NotImplementedError}. Collectively, these tasks were designed to test participants' skills in structured data handling, abstraction, and the use of external libraries, such as \texttt{matplotlib}, for visualization.

\subsubsection{\textbf{Debugging Task.}}
Debugging tasks required participants to locate and fix bugs in pre-written code. Each task contained three syntax errors and two logical errors. The first problem presented a faulty calculator implementation, while the second involved a string utilities module. These tasks included common mistakes such as incorrect operators, misplaced parentheses, or misused functions, and were designed to assess participants' ability to understand and correct faulty code.

\subsection{Metrics}
We measured both objective and subjective outcomes:
\begin{itemize}
  \item \emph{Completion time} and \emph{accuracy} for each task.
  \item \emph{Usability}: System Usability Scale (SUS) \cite{brooke1996sus} scores for each task.
  \item \emph{Workload}: The NASA Task Load Index (NASA-TLX) \cite{hart1988nasatlx} scores for each task.
  \item \emph{Perceived Usefulness (PU)}: Participants' subjective judgment of how much the system improves their performance for each \lipcoder feature.
  \item \emph{Perceived Ease of Use (PEOU)}: The degree to which the participants found the system intuitive and easy to operate for each \lipcoder feature.
  \item \emph{Behavioral Intention (BI)}: The self-reported likelihood of the participants to adopt the system in future use for each \lipcoder feature.
  \item \emph{Qualitative feedback}: Analysis of interview transcripts.
\end{itemize}

We focus on these four metrics because they jointly capture both \emph{objective performance} (speed and effort) and \emph{subjective experience} (workload and usability).
The \emph{System Usability Scale (SUS)} is a widely adopted ten-item questionnaire that provides a quick and reliable measure of general usability, producing a single score that can be compared between systems.
The \emph{NASA Task Load Index (NASA-TLX)} is a multidimensional rating method that captures perceived workload along six factors (mental demand, physical demand, temporal demand, performance, effort, and frustration), offering a sensitive assessment of how taxing an interface feels to users.
This combination of measures is standard in HCI and accessibility research, balancing efficiency-oriented indicators with user-centered assessments.
Accuracy was also recorded; among visually impaired participants it was at ceiling for the boost and debug tasks under both systems, so it discriminates between the systems mainly in the basic task (Tables~\ref{tab:study_outcomes_CL_multirow} and~\ref{tab:sighted_outcomes}).
PU, PEOU, and BI derive from the Technology Acceptance Model (TAM) \cite{dabis1989perceived} and were collected once at the end of the study to assess participants' holistic impressions of \lipcoder. These TAM constructs were reduced to single items for feasibility; they serve as indicative signals of acceptance rather than validated scales.
Together, these instruments complement performance data by highlighting differences in subjective usability and cognitive load between conditions.

\section{Extended Quantitative Results}
\label{app:extended_results}
This appendix elaborates on the statistical analysis and the per-feature perceived acceptance ratings for visually impaired participants reported in Section~\ref{sec:results} (Tables~\ref{tab:study_stats} and~\ref{tab:feature_like}).

\subsection{Significance Testing}
We conducted paired-sample $t$-tests \cite{ttest} or Wilcoxon signed-rank tests \cite{wilcoxon} depending on normality (Shapiro–Wilk test \cite{shapiro}). Familywise error was controlled using Holm–Bonferroni correction \cite{holmbonferroni}, and effect sizes were reported as Cohen's $d_{z}$ \cite{cohen1988statistical} with 95\% confidence intervals. After correction, no comparison reached statistical significance ($q > .05$ in all cases); full results are shown in Table~\ref{tab:study_stats}. Given the small sample size ($n = 5$) and the exploratory nature of this study, the observed numerical trends should be regarded as descriptive and interpreted with caution. We report effect sizes for transparency, but they are inherently unstable at this sample size.

\subsection{Perceived Acceptance}
We measured participants' overall acceptance of each \lipcoder feature using the three TAM constructs:
Perceived Usefulness (PU), Perceived Ease of Use (PEOU), and Behavioral Intention (BI) \cite{dabis1989perceived}; per-feature ratings are shown in Table~\ref{tab:feature_like}.
As described in Table~\ref{tab:feature_mapping}, the features span different modules and programming aspects; here we summarize the most highly rated ones.

Visually impaired programmers consistently assigned the maximum rating (7.0) to multiple features.
For \emph{PU}, they highlighted \textit{settings and help} (F17), \textit{continuous suggestions} (F16), and \textit{code generation and modification} (F10).
For \emph{PEOU}, top scores were given to \textit{settings and help} (F17), \textit{continuous suggestions} (F16), and \textit{file explanation} (F6).
Finally, for \emph{BI}, adoption intentions clustered around \textit{code explanation} (F11),
\textit{code execution} (F12), \textit{pause-triggered completion suggestions} (F15), and \textit{continuous suggestions} (F16).
The consistently high acceptance scores among visually impaired participants underscore the strong perceived value of \lipcoder in supporting accessible programming workflows. Acceptance ratings from sighted participants are reported in Appendix~\ref{app:sighted_study}.

\section{Per-Participant Qualitative Findings}
\label{app:per_participant}
This appendix reports the full qualitative analysis for each visually impaired participant, complementing the cross-participant synthesis in Section~\ref{sec:qualitative}.

\subsection{Participant 1 (P1)}
P1 has severe low vision—she can perceive broad layout changes (e.g., editor switches, large text insertions) but cannot read on-screen text. She had little prior experience with LLM-assisted coding.

Because of her restricted access to textual information, auditory-first features were not supplementary but foundational to her ability to complete tasks. Her workflow followed a ``listen–act–validate'' loop: listening to \lipcoder's spoken summaries, making targeted edits or navigations by voice, and then validating through audio cues rather than scanning code visually. She rarely used keyboard shortcuts, depending almost entirely on ASR commands.

P1 praised voice-based keyword differentiation, explaining that tonal variations reduced fatigue when scanning through long passages of code. Unlike other participants who found multiple voices distracting, she found them clarifying. She also appreciated bracket earcons, though she noted that nested delimiters could sometimes become confusing, suggesting that simpler schemes might be preferable.

Among all features, P1 was most enthusiastic about summarization tools. Both error summarization and file-level code summarization were described as ``essential,'' cutting perceived coding effort by ``tenfold.'' Navigation supports (function lists, symbol trees, natural language commands) were rated not merely as convenient but as necessary to make programming feasible without visual reference. She gave perfect ratings to code modification features, describing them as transformative for allowing her to focus on higher-level intent.

When comparing \lipcoder with Copilot, she highlighted that \lipcoder delivered feedback in concise speech rather than long text passages, making it possible to navigate and validate results without overwhelming memory load. At the same time, she noted that very long outputs would still be better handled through text, suggesting that hybrid modes remain important.

\subsection{Participant 2 (P2)}
P2 is fully blind with extensive experience using LLMs and vibe coding toolkits (Claude, ChatGPT, Windsurf \cite{windsurf}, Gemini-cli \cite{gemini-cli}). He treats LLMs as integral collaborators rather than occasional assistants, engaging in longer, strategic prompts and carefully negotiating context and constraints.

He considered keyword voice differentiation to be of limited utility, since blind developers often memorize reserved keywords and rely on execution feedback, making tonal shifts redundant and sometimes distracting.
By contrast, delimiter earcons were highly valued—distinct rising and falling tones for opening and closing brackets enabled faster error detection than conventional screen reader narration.
Similarly, auditory feedback for indentation was regarded as essential in Python, where whitespace conveys semantic meaning.

Error summarization and syntax alerts received strong endorsements; immediate auditory signals provided a more fluid workflow compared to verbose screen reader feedback. Navigation features, especially natural language commands for opening files and traversing function or class lists, were described as ``game-changing.'' Speech-based code generation and refactoring were described as indispensable, and he also emphasized that image captioning support was extremely valuable for tasks involving visual assets such as charts.

When comparing \lipcoder to Copilot, P2 noted that Copilot required constant text-based interaction, whereas \lipcoder allowed continuous, hands-free coding through integrated auditory feedback. He expressed a desire for compatibility between the two systems, suggesting that combining Copilot's generative power with \lipcoder's auditory-first interface would be particularly effective. His feedback underscores the importance of designing environments that complement, rather than replicate, conventional accessibility tools.

\subsection{Participant 3 (P3)}
P3 has severe low vision with over 20 years of programming experience. He is the only participant who can still read on-screen text—though only by leaning extremely close to the monitor. He described his information flow as approximately 70\% visual (magnified text) and 30\% auditory (screen reader for verification), positioning him between fully auditory-based and visually dominant workflows. He was the most eager to experiment with \lipcoder's full range of features, actively toggling between modalities.

He reacted negatively to keyword voice differentiation, finding timbre changes distracting and cognitively taxing, and recommended alternatives such as pitch shifts or dedicated earcons. By contrast, bracket earcons received perfect ratings across all dimensions—he emphasized that distinguishing opening and closing brackets with distinctive sounds reduced errors, and even suggested expanding this to distinct instrument-like tones for nested structures.

Error summarization was rated highly for debugging, while file-level summarization was seen as valuable but underutilized—he had not noticed the feature initially, stressing the need for built-in tutorials. Natural-language navigation was considered promising, though he again emphasized that discoverability and onboarding tutorials were crucial for adoption.

For code generation, P3 offered some of the most positive evaluations among all participants, reporting that speech-driven generation felt ``lighter on cognitive load'' compared to typing prompts into Copilot. He argued that speech was well-suited for higher-level requests (e.g., ``generate a function for parsing''), while text remained preferable for fine-grained editing or naming identifiers. He stressed that auditory feedback must be carefully designed for clarity and cognitive comfort, and that structured tutorials are critical for enabling experienced developers to adopt new interaction styles.

\subsection{Participant 4 (P4)}
P4 is fully blind and uses VS Code + Copilot + VoiceOver daily—directly corresponding to the baseline environment. His process reflects a pure auditory-first coding paradigm, relying entirely on auditory and keyboard-driven interaction.

He was skeptical of keyword voice differentiation, finding multiple TTS voices distracting. In contrast, he was enthusiastic about panning cues and musical earcons for indentation, describing them as ``much more direct and less error-prone'' than counting spaces or tabs. File summarization and navigation aids (symbol tree, function lists) were praised for dealing with large projects where screen reader navigation becomes cumbersome.

Spoken code generation was described as ``overwhelmingly positive''—voice-based prompting was faster than typing, and errors could be quickly undone with a single command. However, he acknowledged challenges in collaborative settings, expressing concern about how voice-based coding might scale in shared workplaces.

When comparing \lipcoder with his baseline, P4 noted that \lipcoder offered unique strengths in auditory structural feedback and voice-based command execution. However, Copilot with VoiceOver retained advantages for fine-grained review of AI suggestions, since screen reader navigation allows flexible back-and-forth through generated text. He concluded that practical adoption would depend on latency and seamless integration of both voice and keyboard shortcuts, describing \lipcoder as a ``tool built with us in mind.''

\subsection{Participant 5 (P5)}
P5 is fully blind with a decade of professional experience, primarily in Objective-C. His workplace prohibits LLMs and vibe coding toolkits due to regulatory constraints, giving him very limited prior exposure to LLM-assisted workflows. Despite this, he adapted quickly to the experimental environment.

He was especially enthusiastic about summarization tools—spoken summaries of error logs greatly reduced time spent scrolling through text. Hierarchical overviews of project structure and function lists were described as ``indispensable.'' In contrast, he rated keyword differentiation and bracket earcons as low in usefulness, arguing that they duplicated information already provided by screen readers. For indentation, however, he found musical pitch cues promising, emphasizing that hearing indentation levels as rising or falling tones could be far less tiring than counting spaces.

He responded positively to natural-language navigation, noting that voice-driven navigation provided a more fluid alternative to Xcode's rotor-based movement. For code generation and refactoring, he expressed cautious optimism—he valued the potential efficiency but voiced concerns about relinquishing control, worrying that AI-driven transformations might restructure code in undesirable ways. He strongly preferred auditory-first systems over Copilot for multi-file code generation, but emphasized that long prompts remain cumbersome to dictate, suggesting that hybrid input (voice + text) is essential. He underscored that adoption in professional contexts would require configurability—the ability to toggle auditory cues, adjust verbosity, and combine voice with typing depending on the task.

\subsection{Cross-Participant Patterns}
Across participants, auditory navigation and summarization were consistently relied upon to reduce the overhead of line-by-line traversal. Spoken code generation was widely adopted, though at varying granularity—from P4's function-by-function approach to P5's full-file requests. Keyword voice differentiation was the most divisive feature: P1 found it helpful, while P2–P5 found it redundant or distracting. Delimiter earcons and indentation cues were broadly valued, with the exception of P5. Differences also appeared in reliance on ASR versus manual orientation: P1 avoided shortcuts almost entirely in favor of ASR, while P3 still engaged with visual strategies where possible. Collectively, these patterns show that while auditory-first workflows form the foundation, individual strategies adapt based on prior experience, residual vision, and preferred balance between automation and manual validation.

\section{Sighted Participant Study}
\label{app:sighted_study}
This appendix reports in full the sighted comparison study summarized in Sections~\ref{sec:results} and~\ref{sec:qualitative}, run with the same experimental protocol as the visually impaired programmer study.

\subsection{Participants}
A group of 20 sighted programmers (11 male, 9 female; age 21--31; M=3.7 years of Python experience) was recruited. The sighted participants were fluent in Python with at least 2 years of experience.

\subsection{Quantitative Results}

\begin{table*}[t]
  \centering
  \setlength{\tabcolsep}{3pt}
  \renewcommand{\arraystretch}{1.1}
  \caption{Outcomes by task condition for sighted participants (\(n=20\)).
  Each metric is reported as ``\emph{mean} ± \emph{standard deviation}''.
  Boldface numbers indicate better results.
  SUS: System Usability Scale $\uparrow$ (0–100), TLX: NASA Task Load Index $\downarrow$ (0–100), Time $\downarrow$ (min), and
  Accuracy $\uparrow$ (\%).}
  \label{tab:sighted_outcomes}
  \begin{tabular}{l l c *{4}{c}}
    \toprule
    \multirow{2}{*}[-2pt]{\textbf{Task}} & \multirow{2}{*}[-2pt]{\textbf{System}} & &
      \multicolumn{4}{c}{\textbf{Sighted} (\(n=20\))} \\
    \cmidrule{4-7}
    & & & SUS $\uparrow$ & TLX $\downarrow$ & Time $\downarrow$ & Acc. $\uparrow$ \\
    \midrule
    \multirow{2}{*}{\textbf{Basic}}
      & Baseline &   & \resultnum{57.6}{14.4} & \resultnum{45.4}{12.1} & \textbf{\resultnum{14.2}{2.2}} & \resultnum{57.0}{33.3}\\
      & \shade \lipcoder & \shade & \shade \textbf{\resultnum{63.5}{13.4}} & \shade \textbf{\resultnum{41.2}{14.4}} & \shade \textbf{\resultnum{14.2}{1.3}} & \shade \textbf{\resultnum{82.0}{28.2}} \\
    \midrule
    \multirow{2}{*}{\textbf{Boost}}
      & Baseline &   & \resultnum{62.6}{16.5} & \resultnum{41.4}{15.8} & \textbf{\resultnum{12.7}{3.6}} & \textbf{\resultnum{75.0}{33.6}} \\
      & \shade \lipcoder & \shade & \shade \textbf{\resultnum{65.3}{14.7}} & \shade \textbf{\resultnum{40.9}{14.6}} & \shade \resultnum{14.0}{2.3} & \shade \resultnum{71.0}{29.4} \\
    \midrule
    \multirow{2}{*}{\textbf{Debug}}
      & Baseline &   & \resultnum{67.8}{14.5} & \resultnum{34.7}{13.8} & \resultnum{12.2}{4.2} & \textbf{\resultnum{91.0}{10.2}} \\
      & \shade \lipcoder & \shade & \shade \textbf{\resultnum{71.2}{12.3}} & \shade \textbf{\resultnum{34.4}{13.1}} & \shade \textbf{\resultnum{11.4}{4.9}} & \shade \textbf{\resultnum{91.0}{10.2}} \\
    \midrule
    \multirow{2}{*}{\textbf{Overall}}
      & Baseline &   & \resultnum{62.7}{15.5} & \resultnum{40.5}{14.5} & \textbf{\resultnum{13.1}{3.5}} & \resultnum{74.3}{30.8} \\
      & \shade \lipcoder & \shade & \shade \textbf{\resultnum{66.7}{13.7}} & \shade \textbf{\resultnum{38.8}{14.1}} & \shade \resultnum{13.2}{3.4} & \shade \textbf{\resultnum{81.3}{25.2}} \\
    \bottomrule
  \end{tabular}
\end{table*}

\begin{table*}[t]
  \centering
  \setlength{\tabcolsep}{3pt}
  \renewcommand{\arraystretch}{1.1}
  \caption{Statistical significance test between the baseline and \lipcoder for sighted participants (\(n=20\)). Reported are $p$-values (paired t-test or Wilcoxon as appropriate), FDR-corrected $q$-values, and effect sizes (Cohen's $d_z$). Shaded results indicate that \lipcoder showed better results than the baseline.}
  \label{tab:sighted_stats}
  \begin{tabular}{
    l c
    *{3}{c} c
    *{3}{c} c
    *{3}{c} c
    *{3}{c}
  }
    \toprule
    \multirow{2}{*}[-2pt]{\textbf{Task}} & &
      \multicolumn{3}{c}{\textbf{SUS}} & &
      \multicolumn{3}{c}{\textbf{NASA--TLX}} & &
      \multicolumn{3}{c}{\textbf{Time}} & &
      \multicolumn{3}{c}{\textbf{Accuracy}}  \\
    \cmidrule{3-5}\cmidrule{7-9}\cmidrule{11-13}\cmidrule{15-17}
    & & \textbf{p} & \textbf{q} & \boldmath$d_z$ & &
            \textbf{p} & \textbf{q} & \boldmath$d_z$ & &
            \textbf{p} & \textbf{q} & \boldmath$d_z$ & &
            \textbf{p} & \textbf{q} & \boldmath$d_z$ \\
    \midrule
    \textbf{Basic}
     &
       & \shade 0.145 & \shade 0.419 & \shade 0.38
       & & \shade 0.350 & \shade 0.579 & \shade -0.26
       & & \shade 0.388 & \shade 0.600 & \shade -0.01
       & & \shade 0.042 & \shade 0.273 & \shade 0.63 \\
    \midrule
    \textbf{Boost}
     &
       & \shade 0.420 & \shade 0.611 & \shade 0.18
       & & \shade 0.794 & \shade 0.886 & \shade -0.00
       & & 0.192 & 0.419 & 0.28
       & & 0.412 & 0.611 & -0.11 \\
    \midrule
    \textbf{Debug}
     &
       & \shade 0.302 & \shade 0.537 & \shade 0.28
       & & \shade 0.513 & \shade 0.704 & \shade -0.02
       & & \shade 0.535 & \shade 0.714 & \shade -0.14
       & & 1.000 & 1.000 & 0.00 \\
    \midrule
    \textbf{Overall}
     &
       & \shade 0.245 & \shade 0.491 & \shade 0.34
       & & \shade 0.701 & \shade 0.838 & \shade -0.17
       & & 0.793 & 0.886 & 0.06
       & & \shade 0.190 & \shade 0.419 & \shade 0.38 \\
    \bottomrule
  \end{tabular}
\end{table*}

For sighted participants, SUS ratings were numerically higher with \lipcoder (Overall: $62.7 \pm 15.5$ $\rightarrow$ $66.7 \pm 13.7$), and NASA-TLX workload scores were numerically lower (Overall: $40.5 \pm 14.5$ $\rightarrow$ $38.8 \pm 14.1$); full outcomes are shown in Table~\ref{tab:sighted_outcomes}. After correction, no comparison reached statistical significance (Table~\ref{tab:sighted_stats}).

\paragraph{Perceived Acceptance.}
Per-feature acceptance ratings for sighted participants are shown in Table~\ref{tab:sighted_feature_like}.
For sighted programmers, \emph{PU} ratings were highest for \textit{continuous suggestions} (F16),
followed by \textit{code generation and modification} (F10), and \textit{code explanation} (F11).
In terms of \emph{PEOU}, they emphasized the intuitiveness of \textit{delimiter earcons} (F2),
\textit{natural-language code navigation} (F8), and \textit{diverse timbre vocalization} (F1).
For \emph{BI}, the strongest adoption intentions were again for \textit{continuous suggestions} (F16),
alongside \textit{code generation and modification} (F10) and \textit{code explanation} (F11).
Overall, sighted participants prioritized LLM-driven assistance in usefulness and adoption intent, while rating the auditory cues themselves as easy to interpret.

\begin{table*}[t]
\setlength{\tabcolsep}{2.5pt}
\renewcommand{\arraystretch}{1.2}
\caption{Perceived acceptance measures across \lipcoder features for sighted participants (\(n=20\)).
  Each metric is reported on a 7-point Likert scale (1 = strongly disagree, 7 = strongly agree).
  PU: Perceived Usefulness $\uparrow$, PEOU: Perceived Ease of Use $\uparrow$, BI: Behavioral Intention to use $\uparrow$.}
\label{tab:sighted_feature_like}
\definecolor{BlueBase}{HTML}{97C0F6}
\definecolor{RedBase}{HTML}{EB5757}
\begin{tabular}{clccccccccccccccccc}
\toprule
& Metric & F1 & F2 & F3 & F4 & F5 & F6 & F7 & F8 & F9 & F10 & F11 & F12 & F13 & F14 & F15 & F16 & F17 \\
\hline

\multirow{3}{*}{\rotatebox{90}{Sighted}} & PU $\uparrow$ & \cellcolor{RedBase!53} 2.40 & \cellcolor{RedBase!13} 3.60 & \cellcolor{RedBase!63} 2.10 & \cellcolor{RedBase!26} 3.21 & \cellcolor{BlueBase!23} 4.68 & \cellcolor{RedBase!7} 3.79 & \cellcolor{BlueBase!9} 4.26 & \cellcolor{BlueBase!37} 5.11 & \cellcolor{BlueBase!5} 4.16 & \cellcolor{BlueBase!51} 5.53 & \cellcolor{BlueBase!47} 5.42 & \cellcolor{BlueBase!15} 4.45 & \cellcolor{BlueBase!26} 4.79 & \cellcolor{BlueBase!13} 4.40 & \cellcolor{BlueBase!4} 4.11 & \cellcolor{BlueBase!65} 5.95 & \cellcolor{BlueBase!26} 4.79 \\
 & PEOU $\uparrow$ & \cellcolor{BlueBase!80} 6.40 & \cellcolor{BlueBase!87} 6.60 & \cellcolor{BlueBase!68} 6.05 & \cellcolor{BlueBase!75} 6.26 & \cellcolor{BlueBase!75} 6.26 & \cellcolor{BlueBase!68} 6.05 & \cellcolor{BlueBase!70} 6.11 & \cellcolor{BlueBase!81} 6.42 & \cellcolor{BlueBase!77} 6.32 & \cellcolor{BlueBase!75} 6.26 & \cellcolor{BlueBase!79} 6.37 & \cellcolor{BlueBase!73} 6.20 & \cellcolor{BlueBase!75} 6.26 & \cellcolor{BlueBase!67} 6.00 & \cellcolor{BlueBase!44} 5.33 & \cellcolor{BlueBase!79} 6.37 & \cellcolor{BlueBase!75} 6.26 \\
 & BI $\uparrow$ & \cellcolor{RedBase!45} 2.65 & 4.00 & \cellcolor{RedBase!50} 2.50 & \cellcolor{RedBase!4} 3.89 & \cellcolor{BlueBase!30} 4.89 & \cellcolor{RedBase!4} 3.89 & \cellcolor{BlueBase!19} 4.58 & \cellcolor{BlueBase!42} 5.26 & \cellcolor{RedBase!9} 3.74 & \cellcolor{BlueBase!61} 5.84 & \cellcolor{BlueBase!58} 5.74 & \cellcolor{BlueBase!13} 4.40 & \cellcolor{BlueBase!39} 5.16 & \cellcolor{BlueBase!17} 4.50 & \cellcolor{BlueBase!17} 4.50 & \cellcolor{BlueBase!79} 6.37 & \cellcolor{BlueBase!44} 5.32 \\

\bottomrule
\end{tabular}
\end{table*}

\subsection{Qualitative Analysis}

Interviews with sighted programmers revealed nuanced perspectives on the role of audio in programming environments. Overall, participants consistently emphasized that auditory feedback was rarely helpful as a direct replacement for visual information, but they also identified contexts where audio could serve as a complementary channel that enhanced awareness, reduced errors, or supported multitasking. This pattern suggests that while audio features designed primarily for blind users may have limited value for sighted developers, carefully designed auditory augmentations can extend beyond accessibility and provide general productivity benefits.

With respect to auditory output, participants were largely dismissive of functions that attempted to duplicate existing visual distinctions. For instance, gendered voice differentiation or stereo panning of code narration were regarded as distracting or irrelevant, since sighted users already perceive these distinctions quickly on screen. Several participants explicitly stated that ``reading is faster than listening,'' underscoring that redundant auditory presentation of visual data is inefficient. However, they responded more positively to features that acted as error-prevention cues or attention triggers. Auditory icons for parentheses and indentation were seen as helpful reminders that reduced the likelihood of subtle syntactic mistakes, especially when fatigue or inattention might otherwise cause overlooked errors. Some described these cues as making them ``snap back to attention'' during routine editing, suggesting that such earcons serve less as substitutes for vision and more as cognitive nudges that improve vigilance.

Beyond error prevention, some participants also highlighted the affective dimension of auditory feedback. For example, light and pleasant sounds triggered upon certain actions (e.g., pressing Enter) were described as providing a sense of accomplishment or motivation. In this sense, audio was not only a cognitive aid but also an element that enhanced engagement and enjoyment of the programming experience.

Another area where auditory output was judged to have straightforward utility was in summarization and abstraction. Participants frequently expressed that verbatim reading of logs or errors was unhelpful, but concise error summaries or structured overviews of files, variables, and functions could accelerate comprehension. They emphasized that summarization was critical: when audio reproduced line-by-line detail, it became overwhelming; when it abstracted key points, it reduced cognitive load and provided value. This aligns with existing findings that sighted developers prefer higher-level overviews when consuming non-visual feedback. However, participants also stressed that audio summaries lacked the selective navigability of text: whereas written overviews allow quick skipping and scanning, listening required consuming the entire narration. This limitation pointed to the need for hybrid designs that link audio with text-based navigation controls (e.g., ``skip to the second function'').

Multitasking scenarios were also cited as promising contexts for audio. Progress sonification (e.g., TQDM-style auditory progress indicators) was particularly well received, as it allowed programmers to step away from the screen during long-running processes while still monitoring completion status. Several participants noted that they would adopt such features in their own workflows if they were customizable (e.g., adjusting the frequency of updates), highlighting that audio excels when it conveys background information passively without requiring sustained attention. These benefits were especially salient in machine learning contexts, where long model training runs make background auditory updates useful for knowing when to return to the workstation.

In contrast, functions that required continuous focused listening were generally rejected. Stereo panning and gendered speech, for example, demanded conscious effort to interpret and yielded little incremental benefit. Thus, sighted participants distinguished between audio as a primary information channel---which they perceived as slower, redundant, or intrusive---and audio as a secondary, peripheral channel that could provide lightweight awareness and reduce context switching.

Beyond output, participants also reflected on ASR-based input. Here, they identified a double-edged dynamic. On one hand, speech was valued for speed and convenience: when thoughts were clearly formed, dictating commands or requesting code snippets was faster than typing, and participants appreciated that systems often understood even unpolished phrasing. On the other hand, they stressed the difficulty of articulating incomplete or evolving ideas through speech. Textual entry afforded them the ability to iteratively refine, revise, and visualize their thoughts, while speech forced immediate linearization of cognition. Moreover, social and environmental factors---such as the impracticality of talking aloud in shared workplaces---were seen as major constraints on sustained use. Several also noted that, unlike Copilot, which preserves suggestion histories for later inspection, voice-based interfaces provided little transparency or traceability of changes, reducing trust.

Additionally, some participants further underscored that hesitation toward ASR was not only about speed but also about trust and reliability. They voiced concerns about whether the system could truly understand complex textual contexts, explaining that this uncertainty discouraged them from relying on voice input. They suggested that if the system could automatically grasp context and respond confidently, they would prefer voice over text. Others noted that turning on speech recognition, issuing commands, and waiting for responses introduced friction, making them reluctant to use voice in practice. Thus, trust in the system's contextual understanding and reduction of interactional overhead emerged as critical conditions for adoption.

Taken together, these findings indicate that sighted programmers situate audio and ASR features as supplementary modalities rather than replacements for visual programming. They were most receptive to auditory cues that (a) reduce errors by reinforcing critical syntactic events, (b) abstract and summarize complex outputs into digestible units, and (c) provide background awareness during multitasking. They were more cautious about voice input, accepting it for short, command-level interactions but rejecting it as a primary mode of sustained code entry. In addition, the data suggest that audio features may enhance affective engagement (through pleasant or motivational cues), that hybrid text--audio designs are necessary to overcome the linearity of listening, and that trust in ASR's contextual competence is a prerequisite for broader adoption. Thus, for sighted users, the value of audio lies not in replicating visual information but in augmenting visual workflows through attentional cues, lightweight monitoring, and rapid command invocation. These insights suggest that designing auditory features as productivity augmentations---rather than accessibility add-ons---may broaden their adoption and relevance across both blind and sighted developer populations.

\section{Extended Discussion}
\label{app:extended_discussion}
This appendix elaborates on the design lessons and limitations summarized in Section~\ref{sec:discussion}.

\subsection{Design Principles}
Throughout the development of \lipcoder and its evaluation and user feedback, we learned several key lessons for the successful design of a voice-driven programming toolkit.

\subsubsection{\textbf{Unified Yet Modular System.}}
Our experimental comparison of \lipcoder with the baseline system (VSCode + Copilot + VoiceOver) suggested that integrating all core features in a single system may help reduce the cognitive overhead of juggling multiple disparate tools, especially for visually-impaired developers. At the same time, allowing modular customization (substituting and modifying LLM, TTS, and ASR) seemed equally important. Again, this customizability was particularly helpful for visually impaired developers with diverse accessibility needs, such as those who prefer a specific speaker style/timbre of TTS (also see below).

\subsubsection{\textbf{Cross-platform Architecture.}}
During the preliminary investigation, we noticed that each development setup of visually impaired programmers was ``too unique,'' going beyond matching their diverse accessibility needs. We suspected that this was because they rely on several tools that are available only in a particular operating system (OS) or that force unusual workflows tied to one platform, isolating one's setup from others. \lipcoder was hence designed as a VSCode extension with minimal OS dependency. This was important not only to deliver a consistent experience for all users, but also to ensure some standardization of the development setup, while covering the diverse accessibility needs, for easy maintenance and transfer/exchange of development experience between users.

\subsubsection{\textbf{Various User Groups with Different Preferences.}}
Our study revealed diversity among visually impaired programmers in preference for the speech synthesizer. For example, visually impaired programmers preferred the speed and consistency of concatenative speech synthesis; they found neural TTS voice less intelligible, due to its content-dependent prosody, as if the text is ``constantly changing its font'' when reading code. A supplementary study with sighted developers (Appendix~\ref{app:sighted_study}) further confirmed that users for whom audio is primary versus supplementary have different priorities and preferences. Therefore, a voice-driven programming toolkit should offer multiple options to accommodate the diverse habits and needs of different user groups.

\subsubsection{\textbf{Leveraging the Strengths of Auditory Modality.}}
Auditory interfaces offer unique channels to convey information, if used in a thoughtful way. \lipcoder introduced spatialized audio and earcons to represent the structure and events of the code. Participants reported that these cues can aid comprehension, though their perceived effectiveness varied from user to user. For instance, the panning technique to indicate cursor location or indentation was appreciated by low-vision users who could mentally map audio to a visual layout. However, it did not provide much benefit to the participants who were completely blind from birth (P3). Meanwhile, none of our visually impaired participants had prior exposure to earcons in coding, yet after a short learning curve, they reported that earcons were ``highly useful'' and wished to continue using them. Again, while incorporating non-speech audio cues is helpful to convey program state or feedback, they must be configurable and optional.

\subsubsection{\textbf{Speech Input for Efficient Navigation.}}
Voice input can streamline tasks that are cumbersome with screen readers alone. In our study, participants used speech queries (combined with a code context model) to navigate code tasks; rather than listening sequentially through a long file or memorizing complex keyboard shortcuts, a user was able to say, ``Go to the \texttt{renderUserProfile} function'' or ask a question about code functionality, and \lipcoder would jump or summarize accordingly. Participants reported that this was especially helpful for visually impaired developers who are less experienced with advanced keyboard shortcuts.

\subsubsection{\textbf{Ambiguity in Voice Commands.}}
Voice interactions are inherently brief and contextual, which can lead to ambiguity. When a programmer asks, for instance, ``Explain the training function,'' do they mean explaining a specific function named \texttt{train()} in the code, or a general explanation of a training routine? Such vagueness was encountered in our study's conversational coding (vibe coding) scenarios. \lipcoder currently handles such an underspecified case by prioritizing the code context, e.g., the \texttt{train()} function in the current file, if available. However, future improvements are possible by becoming context-aware (more than a simple scope check), interactively asking a clarifying question if needed, or offering multiple responses to such possible interpretations.

\subsubsection{\textbf{High-level Description for Reduced Workload.}}
We found that concise explanations, summarizing code in one or two sentences, are preferred unless more detail is requested afterwards (hence, a coarse-to-fine approach). Lengthy spoken output can burden users' working memory and attention.

\subsubsection{\textbf{Importance of Tutorial.}}
\lipcoder trained various interactions that users must learn, e.g. voice commands and unfamiliar earcon sounds. The participants in our experiments stressed the importance of having a comprehensive tutorial or training mode for such a system. In our controlled study, we guided users through \lipcoder's features, which mitigated confusion; however, a comparable onboarding experience will be crucial for real-world adoption. Unlike typical IDEs, which rarely ship with tutorials (screen readers being a notable exception), an auditory coding tool should teach first-time users how to work with it effectively. This could include interactive tutorials, documentation with examples of voice commands, and practice exercises.

\subsection{Limitations}
While \lipcoder explores promising directions towards an inclusive development environment through voice, the current implementation has several limitations.

\subsubsection{\textbf{Missing Features.}}
While \lipcoder covers core coding practices through voice, as of now, it does not yet cover all the modern IDE features. Examples include: build automation, testing frameworks, version control (\texttt{diff}, \texttt{merge}, and history view), and collaborative features (live sharing, co-editing, and code reviews). Accessing such advanced features requires inefficient solutions, e.g., screen-reading through the feature options. For further development with extra features, we plan to open-source our implementation.

\subsubsection{\textbf{Dependency on Speech and Language Models.}}
The \lipcoder's performance largely depends on the reliability of its components. For example, in noisy environments or with users who have accents that were not adequately represented during training, the current ASR model may incorrectly transcribe their commands. The wrong commands would then be passed to the LLM, which could further cascade the errors by misinterpreting the user's intent. This would eventually lead to frustration and reduced productivity.
Additionally, if the inference and network latencies of the ASR and LLM modules are high, it could bottleneck and interrupt the flow of coding during voice interactions. Through the post-survey, some participants suggested that improving the latency of the current system would further enhance the user experience.
Encouragingly, the performance of ASR and LLM and their inference latency are getting improved quickly; such advanced models should be integrated in the future updates.

\subsubsection{\textbf{Scope of the User Study.}}
While our experiments were carefully designed with details, including multiple tests, quantitative evaluation, and qualitative surveys, they involved a relatively small number of participants, resulting in our evaluation being exploratory in nature.
As a result, no comparison reached statistical significance after multiple-comparison correction.
Furthermore, even with the five participants, it was evident that visually-impaired programmers are diverse in terms of skill level, coding style, and familiarity with assistive technology. Therefore, further in-depth studies are required to examine how well \lipcoder supports all such variations.
Similarly, we have not yet studied group dynamics, such as how a blind programmer using \lipcoder would collaborate with others, including sighted teammates, on a joint project.
Longitudinal studies, observing how developers use \lipcoder over extended periods, would also provide insights into when and how the voice-centric programming toolkits could be useful.

\subsubsection{\textbf{Environmental and Social Factors.}}
\lipcoder assumes that the users can speak out loud and listen to the audio feedback, which may not always be feasible. In a quiet office, library, or classroom, verbally interacting with a coding assistant can be disruptive to others or may not be socially acceptable for the user.
  \fi
\else
  
  \balance                            
  \bibliography{refs}
\fi


\end{document}